\documentclass[%
 reprint,
 amsmath,amssymb,
 aps,
]{revtex4-2}

\usepackage{graphicx}
\usepackage{dcolumn}
\usepackage{bm}
\usepackage{algorithm}
\usepackage{algpseudocode}

\usepackage{braket} 

\usepackage{hyperref}
\usepackage{array}
\usepackage{appendix} 

\begin{document}

\title{\textbf{Quantum block Krylov subspace projector algorithm for computing low-lying eigenenergies}}%

\author{Maria Gabriela Jordão Oliveira}
\email{maria.oliveira@nbi.ku.dk}
\affiliation{%
NNF Quantum Computing Programme, Niels Bohr Institute, University of Copenhagen, Denmark
}
\author{Nina Glaser}
\email{ngl@chem.ku.dk}
\affiliation{%
NNF Quantum Computing Programme, Niels Bohr Institute, University of Copenhagen, Denmark
}%
\affiliation{%
Department of Chemistry, University of Copenhagen, Denmark
}%

\date{\today}

\begin{abstract}

Computing eigenvalues is a computationally intensive task central to many applications in the natural sciences, particularly in quantum physics and chemistry, where determining low-lying energy spectra is key to characterizing physical and chemical properties.
Toward this end, we present and investigate the quantum block Krylov subspace projector (QBKSP) algorithm—a multireference quantum Lanczos method—to accurately compute low-lying eigenenergies, including degenerate ones.
We propose three compact quantum circuits, each suited to different problem settings, for evaluating the necessary expectation values.
In contrast to previous multireference quantum Krylov methods, QBKSP achieves linear scaling with respect to the number of Krylov iterations instead of a quadratic one, and furthermore reduces the scaling with respect to the number of references by half for real Hamiltonians and reference states.
To assess the impact of the number and fidelity of the initial reference states, as well as the QBKSP performance in different parameter regimes, we perform both error-free and sampling noise-limited simulations.
Our results demonstrate that employing multiple reference states significantly improves convergence, particularly in noisy scenarios or when a single reference fails to capture all target eigenstates.
Moreover, the QBKSP algorithm allows for the determination of degenerate eigenstates and their multiplicities through appropriate convergence conditions.

\end{abstract}

\maketitle

\section{Introduction}

The study and prediction of properties of microscopic systems composed of many interacting particles - known as the quantum many-body (QMB) problem \cite{thouless_quantum_1972} - stands as one of the most challenging and computationally expensive tasks in science \cite{feynman_simulating_1982}. 
This problem spans a vast range of fields, including quantum chemistry and molecular, atomic, and condensed matter physics, and solving it enables advancements across various domains in natural sciences. 
However, the complexity of the QMB problem makes the exact calculation of larger systems intractable for classical computers. 
A key problem in this scope is finding eigenenergies of QMB systems, or equivalently, diagonalizing their Hamiltonians. 
To make matrix diagonalization as efficient as possible, more sophisticated classical techniques have been proposed and explored as alternatives to the traditional textbook method. 
Among these techniques are the QR algorithm \cite{francis_qr_1961, francis_qr_1962, kublanovskaya_algorithms_1962}, the Gram iteration \cite{hoffmann_iterative_1989}, the Rayleigh-Ritz procedure \cite{leissa_historical_2005}, and the power method \cite{bai_power_2021}, along with its variants, such as the Arnoldi \cite{arnoldi_principle_1951} and Lanczos \cite{lanczos_iteration_1950, grosso_lanczos-type_1995, gruning_implementation_2011} algorithms.

The growing interest in quantum computation and the hope that, through the exploitation of the underlying quantum mechanical principles, quantum devices will be able to perform some tasks that are prohibitive on classical machines, has led to the development of various quantum and hybrid classical-quantum algorithms. 
These include numerous methods to find eigenvalues, such as quantum phase estimation \cite{kitaev_quantum_1995}, iterative phase estimation \cite{oloan_iterative_2009}, variational quantum eigensolver (VQE) \cite{peruzzo_variational_2014, tilly_variational_2022}, subspace search variational quantum eigensolver (SSVQE) \cite{nakanishi_subspace-search_2019}, quantum multiple eigenvalue Gaussian filtered search (QMEGS) \cite{ding_quantum_2024}, multi-modal multi-level quantum complex exponential least squares (MMQCELS) \cite{ding_simultaneous_2023}, and Krylov subspace diagonalization methods \cite{parrish_quantum_2019, stair_multireference_2020, klymko_real-time_2022, cortes_quantum_2022,baker_block_2024, oumarou_molecular_2025, baker_lanczos_2021, motta_determining_2020, yeter-aydeniz_practical_2020}. 
Some of these quantum algorithms were specifically designed to determine the ground-state energy, while others are capable of retrieving multiple eigenvalues simultaneously, thereby granting access to both ground and excited states.
Although ground state calculations play a crucial role in numerous applications across chemistry and physics, excited states and their corresponding energies are also of great importance. They are, for instance, essential for understanding photochemical reactions, interpreting spectroscopic data, and designing light-emitting devices, among other applications.

Quantum Krylov subspace diagonalization methods, which extend classical Krylov techniques into the quantum domain, are of great interest for extracting multiple eigenvalues on near-term quantum devices.
This interest is motivated by the fact that, as for classical Krylov methods, the size of the Krylov subspace required to retrieve a subset of relevant eigenvalues is expected to be significantly smaller than the original Hilbert space, thus enabling eigenvalue calculations for systems out of reach for exact diagonalization methods \cite{gruning_implementation_2011,grosso_lanczos-type_1995}.
There are various proposals of quantum Krylov algorithms relying on different operators to generate the Krylov subspace; some of them employ the Hamiltonian itself \cite{baker_block_2024,oumarou_molecular_2025}, some the real-time evolution operator \cite{parrish_quantum_2019, stair_multireference_2020, klymko_real-time_2022,cortes_quantum_2022}, i.e., $e^{-i\hat{H}t}$, and others the imaginary-time evolution operator \cite{motta_determining_2020, yeter-aydeniz_practical_2020}, where $t$ is replaced by $-it$, i.e., $e^{-\hat{H}t}$.

As currently available noisy intermediate-scale quantum (NISQ) computers \cite{preskill_quantum_2018} are constrained by the limited qubit counts, the noise present in the devices, and short decoherence times, they are not yet amenable to large-scale error-corrected computations required by more advanced fault-tolerant algorithms. 
Therefore, to harness the potential of near-term quantum hardware, algorithms must be tailored to the current NISQ devices, or to the following generation of devices, i.e., the early fault-tolerant quantum devices \cite{campbell_early_2021, katabarwa_early_2024}, which offer improved noise resilience and larger qubit counts, albeit with some limitations. 

In this work, we investigate the suitability of the quantum block Krylov subspace projector (QBKSP) method for computing ground- and low-lying excited states on early fault-tolerant devices.
The QBKSP algorithm builds on key advances from several prior works, particularly Refs.~\cite{stair_multireference_2020, parrish_quantum_2019, klymko_real-time_2022, cortes_quantum_2022,baker_block_2024}.
Ref.~\cite{parrish_quantum_2019} was the first to propose a real-time evolution-based quantum Krylov subspace method, i.e., the quantum filter diagonalization formalism. 
Ref.~\cite{stair_multireference_2020} firstly introduced a ground-state focused multireference quantum Krylov diagonalization method - the multireference selected quantum Krylov - with the goal of not only avoiding the numerical optimization of parameters as in the traditional VQE but also improving the linear dependency problem of other Krylov methods.
This work also proposes an efficient scheme for the selection of reference states that result in a very compact Krylov basis spanning the ground state wave function.
Independently, Refs.~\cite{klymko_real-time_2022} and \cite{cortes_quantum_2022} proposed real-time evolution-based Krylov variants that advanced techniques for measuring the required quantum expectation values.
Ref.~\cite{baker_block_2024} also presents a variant of multireference quantum Lanczos for excited states relying on the Hamiltonian in itself instead of the real-time evolution to construct the Krylov subspace, but requiring more complex quantum circuits.

To target not only ground but also low-lying excited states, we adopt a multireference framework similar to the algorithm in Ref.~\cite{stair_multireference_2020} in our work, while leveraging the linear time-grid structure introduced in Ref.~\cite{klymko_real-time_2022} for the single-reference case to reduce the number of required quantum circuits.
Since Ref.~\cite{stair_multireference_2020} proposed a reference selection strategy to efficiently span the support space of the ground state which could also be extended to excited states, we assume that suitable initial reference states are available.
To enable a linear scaling of our algorithm with the number of Krylov iterations, we determine the eigenvalues of the time-evolution operator as in Refs.~\cite{klymko_real-time_2022,cortes_quantum_2022} and then map them to the Hamiltonian eigenvalues rather than directly determining the Hamiltonian eigenvalues as in Refs.~\cite{stair_multireference_2020, baker_block_2024}.
By combining the ideas of these works, we present a unified framework that includes not only the multireference Krylov diagonalization in itself but also a regularization routine to prune the errors from quantum hardware and finite state sampling similar to those proposed in Refs.~\cite{epperly_theory_2022,motta_subspace_2024,lee_sampling_2024} and three different compact circuit variations that can be selected depending on the problem and device characteristics to minimize the number of gates and the depth of the circuits.
Moreover, we perform several tests and analyses, including simulations with sampling noise to determine the optimal performance under the realistic constraint of finite shot numbers, thereby gaining new insights into the applicability and performance of the proposed algorithm.

The QBKSP algorithm is outlined in \autoref{sec:algo}, and three different quantum circuits to measure the required expectations values are presented in \autoref{subsec:expval}. 
In \autoref{sec:methods}, we also present the methodology used throughout this work, including the regularization subroutine employed (\autoref{sec:regroutine}), and the setup of all studied models and simulations (\autoref{subsec:modelsimp}).
The QBKSP performance is then analysed in \autoref{subsec:as} on both model spin systems and molecular electronic structure Hamiltonians. 
The studies include three types of experiments: exact numerical simulations (\autoref{subsec:init}, \autoref{sec:molcorr} and \autoref{subsec:time}), simulations with Gaussian sampling noise (\autoref{subsec:precision}) and quantum circuit simulations (\autoref{subsec:quantumsim}). 
Finally, in \autoref{sec:discu}, we discuss the performance of the QBKSP algorithm and different trade-offs that can be made in practical applications.

\section{Algorithm and methodolgy}
\label{sec:methods}
\subsection{Quantum Block Krylov Subspace Projector algorithm}
\label{sec:algo}

The quantum block Krylov subspace projector algorithm is a unification of the single-reference quantum algorithms independently proposed in Refs.~\cite{klymko_real-time_2022} and \cite{cortes_quantum_2022} and the selected multireference method proposed in Ref.~\cite{stair_multireference_2020}.
As such, it is an iterative quantum algorithm for solving the eigenproblem of unitary operators using multiple references, i.e., a block of initial states, instead of just one single reference state.
While physical Hamiltonians are Hermitian, they are not necessarily unitary operators.
To overcome this limitation, the Hamiltonian can be block-encoded \cite{kirby_exact_2023} or expressed through real-time evolution  \cite{stair_multireference_2020, klymko_real-time_2022,cortes_quantum_2022}.
The latter is our method of choice for the QBKSP since it doesn't require any additional ancilla qubits and usually leads to simpler quantum circuits, making the algorithm more amenable to early fault-tolerant devices.
Therefore, assuming atomic units, our aim is to determine the eigenvalues of the real-time propagator $U(t) = e^{-i\hat{H}t}$, for a certain time t, with $\hat{H}$ being the normalized Hamiltonian, i.e., the Hamiltonian rescaled to have its eigenvalues in $[-1,1]$.
The eigenenergies of the Hamiltonian can be retrieved from the phases of the eigenvalues of $U(t)$, by recalling that, for any eigenstate $|\phi\rangle$ of $\hat{H}$,
\begin{equation}
    U(t)|\phi\rangle = e^{-iE_\phi t}|\phi\rangle,
\label{eq:exp}
\end{equation}
where $E_\phi$ is the eigenenergy associated with $|\phi\rangle$. 
Note that, since the phase of $e^{-i\hat{H}t}$ should be in $[-\pi, \pi)$ to resolve the full spectrum, the maximum duration of the time evolution should not be set to values larger than $\pi$ given that the Hamiltonian $H$ is normalized.

Krylov methods enable the determination of a subset of eigenvalues of $U(t)$ by projecting the eigenvalue problem into a smaller subspace, in which a generalized eigenvalue problem
\cite{ghojogh_eigenvalue_2023,parlett_15_1998}
\begin{equation}\label{eq:gevp}
    T|\bm{\phi}\rangle = \bm{\lambda} S|\bm{\phi}\rangle
\end{equation}
can be solved with classical techniques, even for system where diagonalization in the full Hilbert space would be prohibitive.
Here, $T$ contains the expectation values $T_{ij}=\bra{i}U(t)\ket{j}$ and $S$ corresponds to the overlap matrix with elements $S_{ij}=\bra{i}j\rangle$, where $\ket{i}$ and $\ket{j} $ are the elements of the Krylov subspace as defined below.

By definition, a generic order-$r$ Krylov subspace generated by the action of an operator $A$ on a reference state $v$ is given by
\begin{equation}
    \mathcal{K}_r(A, v) = \operatorname{span} \{ v, Av, A^2v, \ldots, A^{r-1}v \}.
\end{equation}

In the QBKSP, we do not only employ a single initial reference, but we start with an initial state block $\{|\psi^{(b)}_0\rangle\}_{b=1}^B$ with $B$ elements.
Based on this initial block, we iteratively build a Krylov basis using the time-evolution operator $U(t)$ with the procedure described in Algorithm \ref{alg:qlanczos}, thus spanning the subspace
\begin{widetext}

\begin{multline}
{\mathcal{K}}_{(K+1)\times B}\left(U(t),\left\{|\psi^{(b)}_0\rangle\right\}_{b=1}^B\right)= \left\{|\psi^{(1)}_0\rangle, |\psi^{(2)}_0\rangle, ..., |\psi^{(B)}_0\rangle, U(t)|\psi^{(1)}_0\rangle, U(t)|\psi^{(2)}_0\rangle, \right. \\ \left. ...,U(t)|\psi^{(B)}_0\rangle, ..., U(t)^{K}|\psi^{(1)}_0\rangle,U(t)^{K}|\psi^{(2)}_0\rangle,...,U(t)^{K}|\psi^{(B)}_0\rangle \right\}.
\label{eq:krylov}
\end{multline}
\end{widetext}
Please note that $U(t)^k = U(t\cdot k)$ for our choice of $U(t)$.

In contrast to the multireference method proposed in Ref.~\cite{stair_multireference_2020}, we do not directly determine the eigenvalues of the Hamiltonian $H$, but instead retrieve them from the phases of the real-time propagator.
Hence, we utilize the same operator $U(t) = e^{-i\hat{H}t}$ to both span the Krylov space and to construct the generalized eigenvalue problem.
A convenient consequence of this choice is that QBKSP only requires a single kind of expectation value to be evaluated, namely $\langle\psi^{\beta}_0| U(t) \ket{\psi^{\alpha}_0}$ at different values of $t$ for $\ket{\psi^{\alpha}_0}, |\psi^{\beta}_0\rangle \in \left\{|\psi^{(b)}_0\rangle\right\}_{b=1}^B$, from which both the overlap matrix $S$ and the $T$ matrix in \autoref{eq:gevp} can be constructed.
As pointed out in Ref.~\cite{klymko_real-time_2022}, by employing a linear time-grid with a fixed time evolution duration $\tau$, we do not need to measure all the elements in the two matrices independently.
In fact, for any $i,j \in \mathbb{N}_0$ and $\ket{\psi^{\alpha}_0}, |\psi^{\beta}_0\rangle \in \left\{|\psi^{(b)}_0\rangle\right\}_{b=1}^B$, the overlaps $\langle \psi^{\beta}_j \ket{\psi^{\alpha}_i}$ simplify to $\langle\psi^{\beta}_0|U^\dagger(\tau)^j U(\tau)^i\ket{\psi^{\alpha}_0} = \langle \psi^{\beta}_0 | U(\tau \cdot (i-j))\ket{\psi^{\alpha}_0}$, and similarly the expectation values $\langle\psi^{\beta}_j| U(\tau) \ket{\psi^{\alpha}_i}$ to $\langle\psi^{\beta}_0|U^\dagger(\tau)^j U(\tau) U(\tau)^i\ket{\psi^{\alpha}_0} = \langle\psi^{\beta}_0| U(\tau \cdot (i-j+1))\ket{\psi^{\alpha}_0}$.
The required expectation values can be measured using the quantum circuits presented in \autoref{subsec:expval}, and the resulting eigenvalue problem can then be diagonalized using classical solvers.

\begin{figure}[htb!]
    \centering 
    \includegraphics[width = 1.0\linewidth]{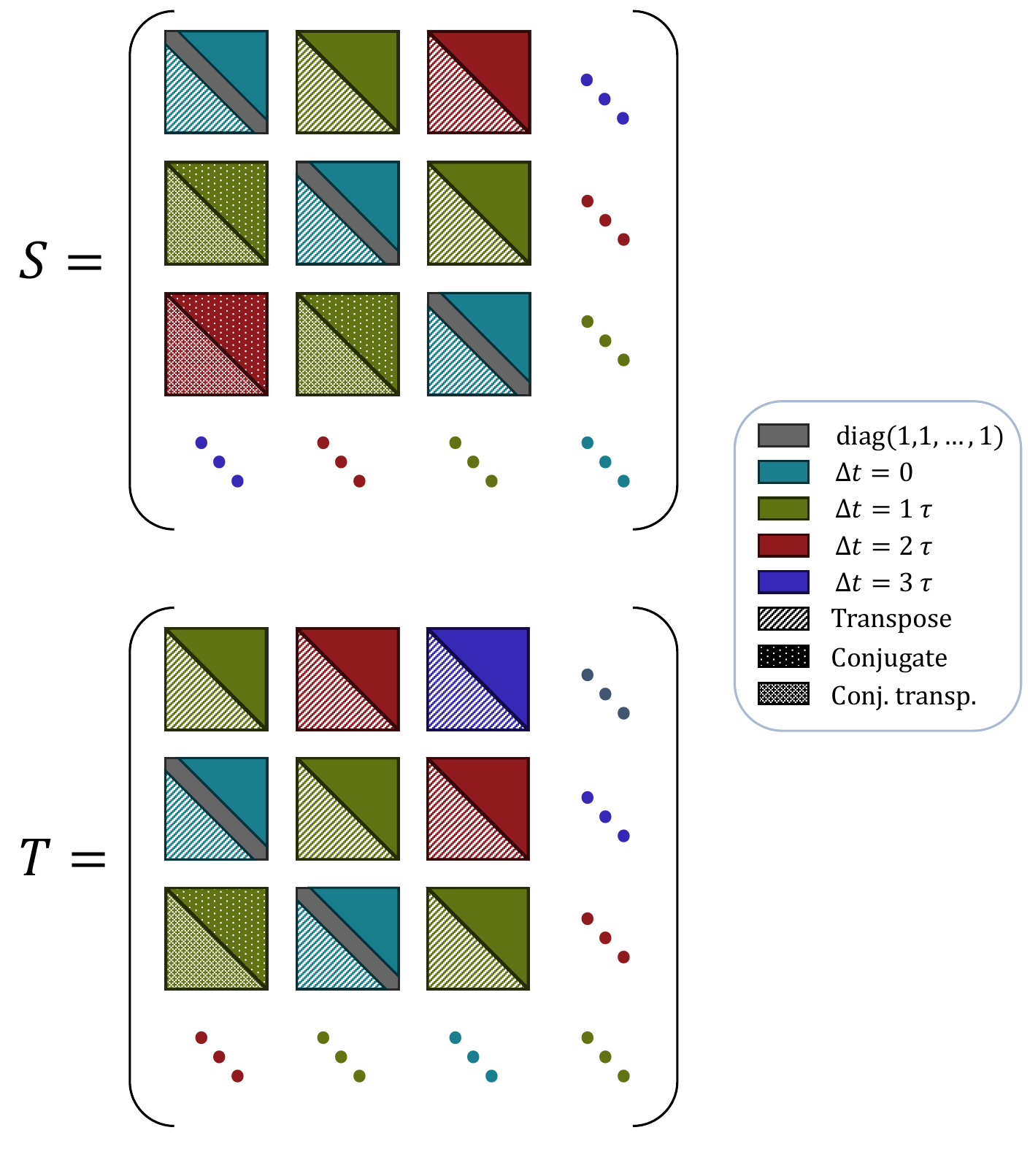}
    \caption{Visualisation of the block-Toeplitz structure of the overlap matrix $S$ and the expectation values in $T$. Each block is of size $B \times B$, where $B$ is the number of reference states, and the different colors represent overlaps of the block Krylov reference states evolved with a given time difference $\Delta t = k \tau$, where $\tau$ is the chosen time step and $k \in {0, 1, ..., K+1}$, with $K$ being the total number of Krylov iterations. The sub-block structure is shown for the case where \( \{|\psi^{(b)}_0\rangle\}_{b=1}^B \) and $\hat{H}$ are real, which results in the individual blocks being symmetric.
    The grey elements denote unit diagonals due to the conservation of the norm under unitary evolution. The different patterns represent transformations of the solid areas of the same color.}
    \label{fig:matrices}
\end{figure}

Leveraging a linear time-grid structure for $U(t)$ also reduces the computational scaling of the algorithm by exploiting the fact that the obtained QBKSP matrices are highly structured, with both $S$ and $T$ resulting in a block-Toeplitz form as shown in \autoref{fig:matrices}.
Furthermore, all elements contained in the overlap matrix $S$ also appear in the $T$ matrix, and consequently the overlap matrix can be constructed by simply recycling the corresponding blocks of the $T$ matrix.
Therefore, for an arbitrary Hamiltonian, the algorithm requires the evaluation of only $B^2 (K+1)+\frac{1}{2}B(B-1)$ distinct measurement values to construct both the $S$ and $T$ matrices. If the initial references are known to be orthogonal, this number is further reduced to $B^2 (K+1)$. 
Regardless of whether the chosen references are orthogonal or not, QBKSP achieves linear scaling with the number of Krylov iterations, resulting in a significant reduction of the computational cost compared to the quadratic scaling of the original multireference algorithm proposed in Ref.~\cite{stair_multireference_2020} which required the evaluation of $B^2(K+1)^2$ expectation values.

Moreover, when the Hamiltonian is real, which is the case for the systems of interest in this work, including the Hubbard, Heisenberg, and molecular electronic structure Hamiltonians, $U(t)$ is complex symmetric.
Under this condition, for real initial reference states the number of expectation evaluations can be reduced from $B^2 (K+1)+\frac{1}{2}B(B-1)$ to  $\frac{1}{2} B (B+1)(K+1) +\frac{1}{2}B(B-1)$, since $\langle \psi_0^{\Re_\alpha}|U(n\cdot\tau)|\psi_0^{\Re_\beta}\rangle = \langle \psi_0^{\Re_\beta}|U(n\cdot\tau)|\psi_0^{\Re_\alpha}\rangle$,  for any $|\psi_0^{\Re_\alpha}\rangle, |\psi_0^{\Re_\beta}\rangle \in \left\{|\psi^{(\Re_b)}_0\rangle\right\}_{b=1}^B$.
In this case, QBKSP further reduces the scaling concerning the number of reference states by a factor of $\frac{1}{2}$ by exploiting the sub-block symmetry of the matrix elements as visualized in \autoref{fig:matrices}.
Independently of the Hamiltonian choice, in the limiting case where only a single reference is employed, QBKSP requires just $(K+1)$ expectation values, essentially corresponding to the single reference algorithm presented by Klymko \textit{et al.}\cite{klymko_real-time_2022} with $(K+2)$ distinct matrix elements, which can be reduced by one when accounting for the unit norm of any physical quantum state.

The fact that the orthogonality of the Krylov basis is not ensured by the algorithm, in addition to the inherent noise of near-term quantum devices and the statistical errors coming from the finite sampling of the matrix elements, can lead to ill-conditioned overlap matrices $S$.
For practical applications of the QBKSP algorithm, the generalized eigenvalue problem should therefore be regularized before solving it classically.
For this purpose, we employ a subspace projection-based regularization scheme in the diagonalization step of QBKSP to prune errors, as described in \autoref{sec:regroutine}.

\begin{algorithm}[H]
\caption{Quantum Block Krylov Subspace Projector (QBKSP) algorithm}
\label{alg:qlanczos}
\begin{algorithmic}[1]
\State \textbf{Input:} Initial states \( \{|\psi^{(b)}_0\rangle\}_{b=1}^B \), unitary operator \( U(\tau) \approx e^{-i\hat{H}\tau} \), target number of eigenvalues  \( N \), maximum number of Krylov iterations \( K \) and convergence criteria
\State Initialize \(B \cdot (K+1) \times B \cdot (K+1)\) complex matrices \(T\) and \(S\)
\For{\(b_i = 1,\dots,B\) and \(b_j = b_i,\dots,B\)} \Comment{Only required for non-orthogonal reference states}
    \If{$b_i = b_j$} $A_{b_i, b_j}^{(0)} \gets 1$
    \Else ~$A_{b_i, b_j}^{(0)} \gets \langle \psi^{(b_i)}_0 | \psi^{(b_j)}_0 \rangle$ \Comment{Evaluate using one of the quantum circuits in \autoref{H circuit}}
    \EndIf
\EndFor
\For{\(k = 0\) to \(K\)} \Comment{Iteratively expand Krylov space}
        \For{\(b_i = 1,\dots,B\) and \(b_j = 1,\dots,B\)}
            \State \( A_{b_i, b_j}^{(i)} \gets \langle \psi^{(b_i)}_0 | U(\tau)^{k+1}| \psi^{(b_j)}_0 \rangle\)
            \Comment{Evaluate using one of the quantum circuits in \autoref{H circuit}}
            \For{\(l = 0\) to \(k\)}
                \State \(I \gets l \cdot B + b_i - 1\), \(J \gets k \cdot B + b_j - 1\) \Comment{Element indices}
                \State \(S[I, J] \gets A_{b_i, b_j}^{(k-l)}\) \Comment{Upper triangular $S$}
                \State \(T[I, J] \gets A_{b_i, b_j}^{(k-l+1)}\)  \Comment{Upper triangular $T$}
                \If{$k\neq l$}
                   \State \(S[J, I] \gets A_{b_i, b_j}^{(k-l)*}\) \Comment{Rest of $S$}
                    \If{$k-l=1$}
                    \State \(T[J, I] \gets A_{b_i, b_j}^{(k-l-1)}\) \Comment{Subdiagonal of $T$}
                    \Else
                    \State \(T[J, I] \gets A_{b_i, b_j}^{(k-l-1)*}\) \Comment{Rest of $T$}
                \EndIf
                \EndIf
            \EndFor
        \EndFor

    \State Solve \(T|\bm{\phi}\rangle = \bm{\lambda} S|\bm{\phi}\rangle\)
    \Comment{Optional: Apply regularization subroutine \ref{alg:regularization}}
    \If{N eigenvalues satisfy convergence criteria}
    \Comment{\textit{E.g.}, $|\lambda_n^{(k)} - \lambda_n^{(k-1)} |\leq \Delta E_n$}
    \State \textbf{break for loop}
    \EndIf
\EndFor
\State \textbf{Output:} Renormalized angles of the eigenvalues \(\bm{\lambda}\)
\end{algorithmic}
\end{algorithm}

The ill-conditioning of the generalized eigenvalue problem can lead to the appearance of artificial degeneracies, as noted in Ref.~\cite{baker_block_2024,epperly_theory_2022}.
Therefore, to accurately determine the correct multiplicities of degenerate states using the QBKSP algorithm, the eigenvalue problem should be regularized and the energy and associated multiplicity of each eigenvalue should be recorded immediately upon its convergence.
In other words, one should start the algorithm by setting a state-specific convergence criterion for each eigenvalue. Then, when the ground state is converged according to the set convergence criterion, its energy and multiplicity should be fixed at this point, and the algorithm continues. Then a similar procedure is employed for the following eigenvalues until the general convergence criteria are satisfied, e.g., $N$ eigenvalues are found according to the individual convergence criteria, with $N$ corresponding to the desired number of eigenvalues to compute.

\subsection{Expectation value measurement}
\label{subsec:expval}
\begin{figure}[htb!]
 \centering 

\includegraphics[width = 1\linewidth]{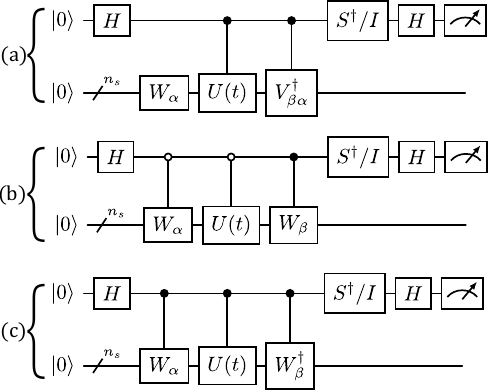}
\caption{Three quantum circuits for retrieving $\bra{\psi^\beta}U(t)\ket{\psi^\alpha}$, with $n_s$ being the number of qubits representing the system of interest.
By replacing the gate $S^\dagger \slash I$ by $I$, the real part of the expectation value is determined, and by setting it to $S^\dagger$, the imaginary part is retrieved. In the first circuit $\ket{\psi^\beta}=V_{\beta\alpha} \ket{\psi^\alpha}$, in the other two  $\ket{\psi^\beta}=W_\beta \ket{\mathbf{0}}$, and in all of them $\ket{\psi^\alpha}=W_\alpha \ket{\mathbf{0}}$. 
For each problem instance, initial reference states, and device characteristics, the quantum circuit version that leads to more efficient quantum circuits should be chosen.
For further details on these three quantum circuits, see Appendix \ref{app:quantumc}.}
    \label{H circuit}
\end{figure}

To construct the generalized eigenvalue problem spanned by the Krylov subspace as given in \autoref{eq:krylov} , the expectation values $\langle\psi^\beta|U(n\tau)|\psi^\alpha\rangle = \langle\mathbf{0}|W^\dagger_\beta U(n\tau) W_\alpha|\mathbf{0}\rangle$ have to be evaluated, where $|\mathbf{0}\rangle$ corresponds to all qubits in state $|0\rangle$. 
Here, the operator $W_\alpha$ when applied to $|\mathbf{0}\rangle$ prepares the initial reference state $|\psi^\alpha\rangle$, and similarly $W_\beta$ when applied to $|\mathbf{0}\rangle$ prepares $|\psi^\beta\rangle$.
The operator $V_{\beta\alpha}$ prepares the state $|\psi^\beta\rangle$ when applied to $|\psi^\alpha\rangle$.
The required expectation values can be evaluated using either one of three quantum circuit variants of the Hadamard test, as shown in \autoref{H circuit}. 
Even though these three quantum circuits allow for the retrieval of the same expectation values, they have different controlled operators, and consequently, they result in different gate counts and circuit depths. 
The optimal version depends on the specific problem, the complexity of the state preparation procedure, and the hardware characteristics. 
For a given choice of circuit type, two different variations of this type are executed for the evaluation of each expectation value: one with the gate $S^\dagger \slash I$ replaced by $I$ to evaluate the real part, and one replaced by $S^\dagger$ to retrieve the imaginary part.

While we present the three quantum circuits in \autoref{H circuit} as different options to measure the required expectation values, there are also more NISQ-friendly alternatives to do so which could be applied within the QBKSP framework \textit{in lieu} of the Hadamard test circuits.
For instance, Refs.~\cite{cortes_quantum_2022,stair23_stochastic} and \cite{yang_phase-sensitive_2024} propose a methodology which reduces the number of controlled time evolutions, or, alternatively, one could also measure the expectation value of the Hamiltonian directly term by term \cite{tilly_variational_2022,stair_multireference_2020,yoshioka_krylov_2025}, as it is typically done in VQE applications.

\subsection{Regularization routine}
\label{sec:regroutine}

To improve the QBKSP convergence in practical applications, we employ a regularization routine to address the numerical instabilities arising from the Lanczos algorithm, as well as errors arising from finite state sampling and the noise present in quantum devices.
As in Refs.~\cite{motta_subspace_2024,klymko_real-time_2022,epperly_theory_2022,lee_sampling_2024}, we regularize the ill-conditioning of the overlap matrix $S$ by introducing a truncation threshold to avoid numerical instabilities.
Specifically, we project both $T$ and $S$ onto the basis of the singular value decomposition of $S$, followed by the removal of the dimensions corresponding to the singular values below a predefined truncation threshold, as described in Algorithm \autoref{alg:regularization}.
Although this routine improves the conditioning of the $S$ matrix, one should be aware that it also introduces a bias in the results as components below the specified truncation threshold are not considered when solving the generalized eigenvalue problem.
Thus, the truncation threshold should be as small as possible while remaining above the noise level of the matrix elements.

\begin{algorithm}[H]
\caption{Noise and sampling error regularization}
\label{alg:regularization}
\begin{algorithmic}[1]
\State \textbf{Input:} Overlap matrix $S$, matrix $T$, threshold parameter $\epsilon$ 
\State Compute the singular value decomposition (SVD) of $S$: $U, \Sigma, V = \text{SVD}(S)$
\State Determine the singular values above the threshold $\epsilon$: $\text{Ids} \gets \{i:\Sigma_i > \epsilon \}$ 

\State Project $T$ and $S$: $\widetilde{T}=U^{-1}TV^{-1}$ and $\widetilde{S}=U^{-1}SV^{-1}$

\State Filter $\widetilde{T}$ and $\widetilde{S}$: $\widetilde{T} \gets \widetilde{T}(\text{Ids},\text{Ids})$ and $\widetilde{S} \gets \widetilde{S}(\text{Ids},\text{Ids})$
\State \textbf{Output:} Compressed matrices $\widetilde{T}$ and $\widetilde{S}$
\end{algorithmic}
\end{algorithm}

\subsection{Models and implementation details}
\label{subsec:modelsimp}
As proof-of-concept applications of the proposed QBKSP algorithm, we apply our methodology to both molecular and model Hamiltonians and analyze the performance of the method for various hyperparameter settings.
As test molecules, we chose lithium hydride and hydrogen fluoride, both modelled using the STO-3g basis set \cite{hehre_self-consistent_1969}. 
The Li-H and H-F bond lengths are set to $1.6$ \AA \ and  $0.91$ \AA, respectively, unless otherwise specified.
As a model system, we employ the Heisenberg \cite{heisenberg_zur_1928} spin chain, whose Hamiltonian is defined as 
\begin{equation}
    \hat{H}_\text{Heisenberg} = -\vec{\mathbf{J}}\sum_{\langle i,j\rangle} \vec{\sigma}_i \otimes \vec{\sigma}_j,
\end{equation}
where the summation is performed over adjacent lattice sites $i$ and $j$.
$\vec{\sigma}_i=\left(\sigma_x, \sigma_y, \sigma_z\right)$ is a vector encompassing the Pauli matrices and $\vec{\mathbf{J}}$ is a vector of coupling constants that are all set to $1$ Hartree in this work.

For all test systems, we first perform numerical simulations under ideal conditions, i.e., in the absence of quantum device noise, Trotterization errors, or statistical errors due to finite sampling effects.
The goal is to study the impact of both the size and the fidelity of the initial states' block as well as the duration of the time evolution.
In addition, we also perform numerical simulations with Gaussian sampling noise added to the expectation values to analyse how the precision of the matrix elements, which is inherently limited by the finite number of shots, affects the results. 
As each measurement of the ancilla qubit corresponds to a binomial event, repeatedly measuring the same quantum circuit yields a probability distribution. 
According to the central limit theorem, in the limit of a large number of shots, the statistical fluctuations of the estimated probabilities converge to a Gaussian distribution around the underlying expectation values \cite{1997zoller}. 
Therefore, employing a Gaussian noise model is a common technique for modeling finite sampling effects in quantum emulations \cite{Oliv2022, filip_fighting_2024}.
These numerical simulations aim to establish the error-free performance baseline of the QBKSP algorithm without any Trotterization errors or quantum device noise, but including finite sampling effects.

For the Heisenberg spin chain, we apply the QBKSP algorithm using an initial-state block composed of 1, 2, and 3 reference states for specific overlaps $\gamma$ with the target states. 
For a given $\gamma$, if we use a single reference state, it has overlap $\gamma$ with the ground state. 
When using 2 (or more) reference states, one of the states has $\gamma$ overlap with the ground state and the other with the first excited state, and so forth.
The initial states are generated randomly under the imposed overlap constraints. 
For molecular systems, we use the Hartree-Fock state as a reference for the ground state, and excited-states guesses are constructed by applying the $\hat{\mu}_x$, $\hat{\mu}_y$, and $\hat{\mu}_z$ dipole operators to the HF state \cite{singh_excitations_nodate,bonaiti_ab_2022}.
Thus, when using only one reference state, we use the HF state; with two reference states, we use the HF state and the HF state acted upon by the $\hat{\mu}_z$ dipole operator; for 3 references, we use the HF state acted on individually by $\hat{\mu}_x$ and $\hat{\mu}_z$ and the HF state itself; and for four references, we include all three dipole-modified variants and the HF state itself.
The choice of the two and three reference states corresponds to the optimal one of this set for the given molecular geometry, as shown in Appendix \ref{app:initmole}.
Alternatively, when no prior knowledge of the system’s eigenstates or dominant occupations is available, the initial reference states can be pre-selected using either classical techniques, for example based on coupled cluster calculations \cite{COESTER19584,CCBartlett}, or also with more approximate quantum computations, such as variants of the reference selection scheme proposed in Ref. \cite{stair_multireference_2020}.
For strongly correlated systems, suitable initial states could, for instance, be obtained from low-cost density matrix renormalization group calculations, as the resulting matrix product states are optimized to capture the pronounced multiconfigurational character of the wavefunction while remaining amenable to efficient preparation on quantum hardware \cite{Malz_2024_MPSprep,Smith_2024_MPSprep}.

For all models, we analyse three evolution times, that is, $\tau=1$, 2 or 3 atu, and we also examine three different sampling noise magnitudes, i.e., Gaussian sampling noise with standard deviations of $\sigma=10^{-3}$, $10^{-6}$ and $10^{-9}$.
In practical applications, the regularization threshold should be adapted to the expected noise level in order to reduce numerical instabilities, and is thus set to $\epsilon=100\sigma$.
Unless mentioned otherwise, by default we utilize an evolution time step of 3 atu and set threshold parameter $\epsilon$ of the $S$ matrix to $10^{-10}$.

In addition to numerical calculations, we also perform quantum circuit simulations of the QBKSP algorithm for the LiH molecule using Qiskit's \textit{Qasm Simulator} \cite{qiskit2024}.
Specifically, we employ a second-order Trotterization scheme \cite{hatano_2005,berry_efficient_2007} to time-evolve the Hamiltonian, and we utilize the quantum circuit b) of \autoref{H circuit} to evaluate the required matrix elements.
We perform simulations for both one reference state (the HF state) and four reference states (the HF state and its three dipole-excited variants) with $10^5$ shots for a Trotter step size of $0.07$ atu.
In these simulations, we set the $S$ matrix threshold parameter to $0.1$ to regularize the eigenvalue problem.

\section{Simulation results and performance analysis}
\label{subsec:as}
In this section, we analyze the performance of the QBKSP algorithm on various test systems, including the Heisenberg model as well as the molecular electronic structure Hamiltonians of lithium hydride and hydrogen fluoride, to demonstrate that QBKSP successfully retrieves low-lying eigenstates both for model Hamiltonians and molecular problems.
We investigate the impact of various hyperparameters on the QBKSP convergence, and study the performance of the method with error-free numerical calculations, as well as in more realistic settings with sampling noise and with quantum circuit simulations.

\subsection{Reference state block dependence}
\label{subsec:init}
\begin{figure*}[htb!]
    \centering
    \includegraphics[width=1\linewidth]{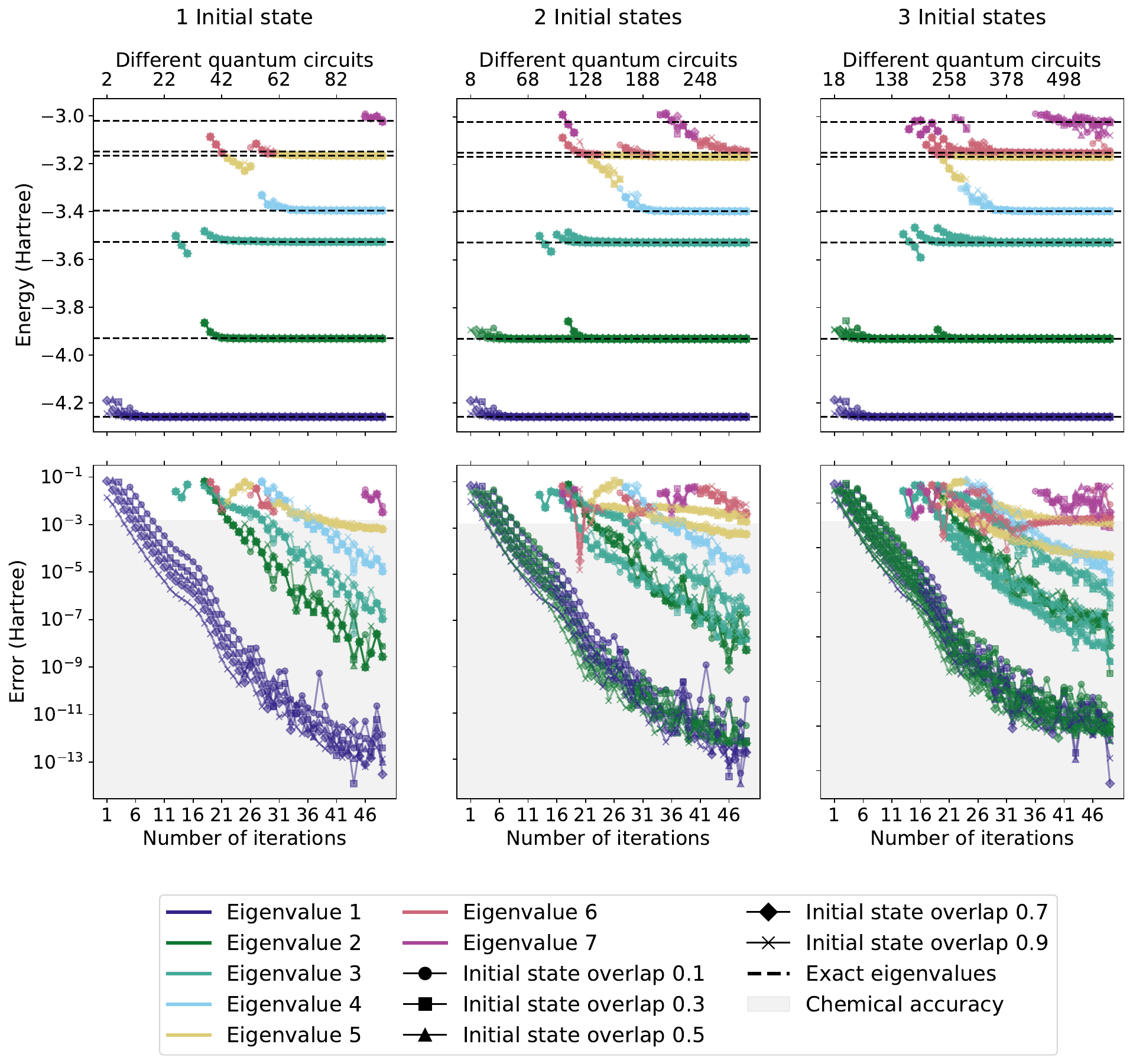}
    \caption{QBKSP convergence for the 10-site Heisenberg model, evaluated for different numbers of initial states and several overlap values.
    The top panel shows the convergence of the seven lowest eigenvalues as a function of both the Krylov iteration  and the number of different quantum circuits required, whereas the bottom panel shows the absolute error with regard to the exact eigenvalues.
    The model contains three 2-fold degeneracies in the seven lowest-lying distinct eigenvalues.
    Note that we define chemical accuracy as errors below $1.6$ mHartree. }
    \label{fig:10Heisenberginitial}
\end{figure*}

To investigate whether the use of multiple initial reference states in QBKSP is advantageous compared to employing only a single-reference state, we compare the performance of QBKSP to the single-reference variant and study the impact of both the number and fidelity of the initial states.
In \autoref{fig:10Heisenberginitial}, we present the convergence behavior for the seven lowest eigenenergies of the 10-site Heisenberg model for one, two, and three reference states, as well as for varying overlaps of the initial states.
The inclusion of multiple reference states enables the retrieval of degenerate states that cannot be resolved when only using a single reference state, as summarized in \autoref{tab:initial}. 
In this case, the use of two and three initial reference states allows for the retrieval of the degeneracies of the first excited eigenstate.

For demonstration purposes, in \autoref{fig:10Heisenberginitial} we let the algorithm run for more iterations than strictly necessary for the lower eigenstates. 
Even though the degenerate eigenenergies are found, their multiplicity is sometimes higher than the correct one since spurious eigenvalues can appear due to numerical errors, such as the ill-condition of the $S$ matrix \cite{baker_block_2024}. 
However, when imposing appropriate convergence criteria as discussed in \autoref{sec:algo}, this issue disappears.
The use of higher threshold parameters for the regularization of the $S$ matrix can also mitigate the overcounting of degenerate states, albeit at the cost of limiting the achievable accuracy as shown in Appendix \ref{app:Sthre}.

\begin{table}[htb!]
    \centering
    \begin{tabular}{|l|l|l|}
    \hline
    Initial ref. states   & Eigenvalues &  \# degeneracies
      \\ \hline \hline
        1 & 5 & 0\\ \hline
        2 & 5 & 1\\ \hline
        3 & 6 & 1\\ \hline
    
    \end{tabular}
    \caption{Number of distinct eigenvalues converged to within chemical accuracy and degeneracies found using QBKSP with $1$, $2$, and $3$ initial reference states, each with $0.5$ overlap, for the 10-site Heisenberg model with 50 Krylov iterations. 
    In the first seven distinct eigenvalues, the model contains three 2-fold degeneracies.
    The data corresponding to these values is shown in \autoref{fig:10Heisenberginitial}.}
    \label{tab:initial}
\end{table}

\begin{figure}[htb!]
    \centering
    \includegraphics[width=0.9\linewidth]{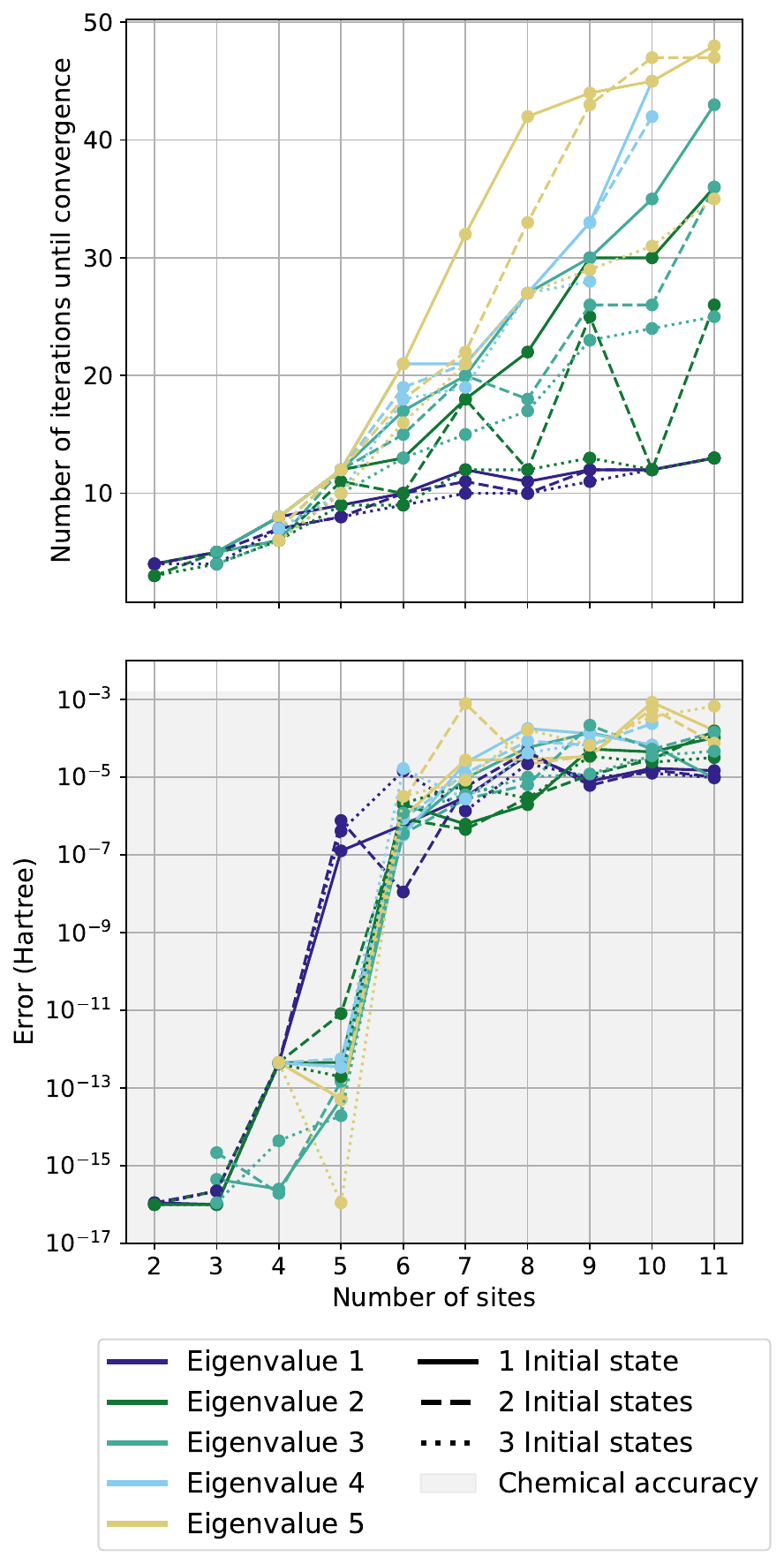}
    \caption{Convergence behavior of the QBKSP algorithm for the Heisenberg model as a function of system size. 
    To reflect a practical application of the method to a previously unsolved use-case, no knowledge about the exact eigenenergies is assumed. To terminate the algorithm, convergence is defined as two consecutive iterations where the calculated energy changes less then $0.1$ mHartree.
    The top panel displays the number of Krylov iterations required to reach convergence, while the bottom panel shows the error of the retrieved energy when compared to the exact energy.
    The convergence of the five lowest (non-degenerate) eigenvalues is shown for one, two, and three initial reference states, each with a fixed overlap of $0.5$.
    Notably, all the retrieved eigenvalues have errors below chemical accuracy.}
    \label{fig:initialconv}
\end{figure}

To further examine how the convergence of the lowest five eigenvalues evolves with the size of the problem, in \autoref{fig:initialconv} we analyze the number of iterations required for increasing lattice sizes of the Heisenberg model. 
The evolution time is fixed at $\tau=3$ atu, and the initial state overlap is set to $0.5$.
The results support the observation that, as for the classical Lanczos algorithm, the Krylov subspace is usually significantly smaller than the original Hilbert space.
Additionally, we find that even without utilizing any knowledge of the true eigenspectrum to check convergence, as is the case for most practical applications, we retrieve well-converged energies for all five states.

Similarly to the analysis in \autoref{tab:initial}, \autoref{tab:initialmol} examines the impact of using different numbers of initial states, under the same fixed evolution time, for the electronic Hamiltonian of lithium hydride.
Through the use of at least three initial reference states, namely the Hartree--Fock (HF) state and the HF state acted upon individually with the dipole operators $\hat{\mu}_x$ and $\hat{\mu}_z$, the QBKSP is able to recover almost the full spectrum of the LiH molecule, improve the accuracy of the computed energies, and with four initial reference states also degenerate eigenstates can be resolved. 
In fact, in this case, all retrieved eigenstates have the correct multiplicity.
The QBKSP convergence for this system is shown in Appendix \ref{app:LiHconv}.
Moreover, in \autoref{tab:supervs4}, we show that using the four just-mentioned initial reference states is also advantageous over starting with a single reference state corresponding to a uniform superposition of these four individual states. Particularly, the use of four initial reference states not only allows the retrieval of degenerate eigenvalues but also requires fewer Krylov iterations and fewer distinct quantum circuits to retrieve the same number of eigenvalues.

\begin{table}[htb!]
    \centering
    \begin{tabular}{|l|l|l|}
    \hline
      Initial ref. states   & Eigenvalues &  \# degeneracies\\ \hline \hline
        1 & 6 & 0\\ \hline
        2 & 7 & 0\\ \hline
        3 & 10 & 0\\ \hline
        4 & 10 & 3\\ \hline
    
    \end{tabular}
    \caption{Number of distinct eigenvalues converged to within chemical accuracy and degeneracies found for the LiH molecule using $1$, $2$, $3$ and $4$ initial reference states at the same computational cost.
    The LiH spectrum contains four 2-fold degeneracies, and the total number of distinct eigenstates is 11.
    The data corresponding to these values is shown in \autoref{fig:LIHinitial}. 
    }
    \label{tab:initialmol}
\end{table}

\begin{table}[htb!]
    \centering
    \begin{tabular}{|l|l|l|l|}
    \hline
      Initial ref. states   & Eigenvalues &  \# degeneracies & \# circuits \\ \hline \hline
        Superposition & 10 & 0 & 172\\ \hline
        HF, $\hat{\mu}_x$, $\hat{\mu}_y$ \& $\hat{\mu}_z$ & 10 & 3 &92\\ \hline

    \end{tabular}
    \caption{Number of distinct eigenvalues converged to within chemical accuracy and degeneracies found for the LiH molecule and respective number of distinct quantum circuits required using both one and four initial state. 
    The single initial state corresponds to a uniform superposition of the HF state and the three excited variants of the HF and four initial reference states correspond to these four states individually.
    The data corresponding to these values is shown in Appendix \ref{app:LiHconv}. 
    The total number of distinct eigenstates in the LiH spectrum is 11 and there are 4 degenerate states.}
    \label{tab:supervs4}
\end{table}

\subsection{Correlation effects in molecular systems}
\label{sec:molcorr}

\begin{figure}[htb!]
    \centering
    \includegraphics[width=0.9\linewidth]{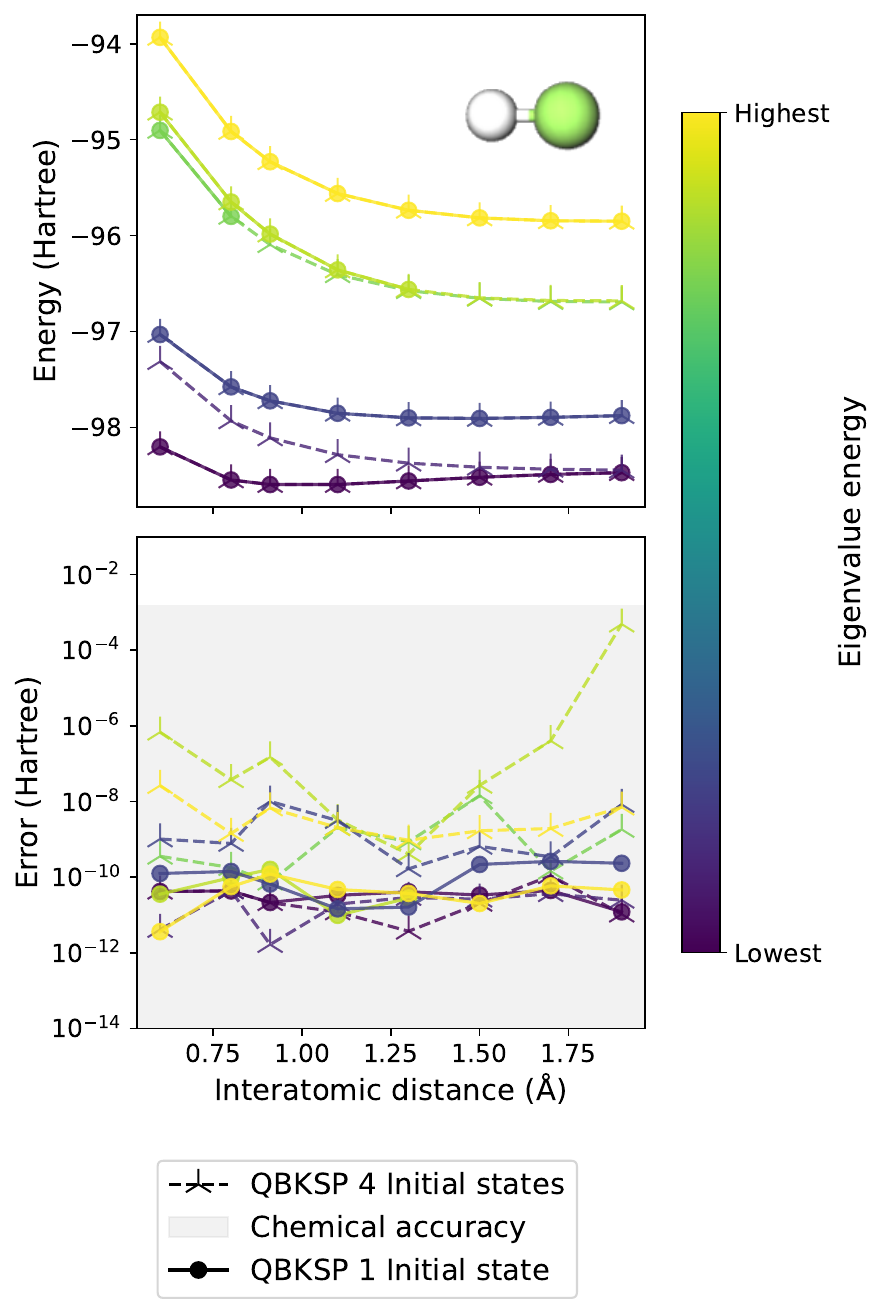}
    \caption{QBKSP algorithm applied to hydrogen fluoride using both one and four initial states.
    To compare the QBKSP performance at the same computational cost, the total number of quantum circuits was fixed to 192, i.e., 96 iterations for the single-reference case, and 8 for the four-reference one.
    The top panel shows the convergence of the three highest and three lowest eigenvalues as a function of the interatomic distance and the bottom panel displays the absolute error with respect to the exact energies.
    }
    \label{fig:HFdist}
\end{figure}

\autoref{fig:HFdist} shows how QBKSP with both the four above-used initial reference states and only the HF state performs in different correlation regimes by analyzing its convergence at different interatomic distances of hydrogen fluoride. 
As the interatomic distance increases, the electronic structure becomes more correlated, making the calculations more challenging at stretched geometries.
Nonetheless, the QBKSP results demonstrate that these four initial states consistently lead to accurate solutions for hydrogen fluoride energy levels across varying bond lengths.
In fact, the results demonstrate that using four reference states enables the computation of eigenenergies that are unreachable using only one reference, even for molecular structures far from equilibrium.

\subsection{Choice of evolution time}
\label{subsec:time}

\begin{figure}[htb!]
    \centering
    \includegraphics[width=0.9\linewidth]{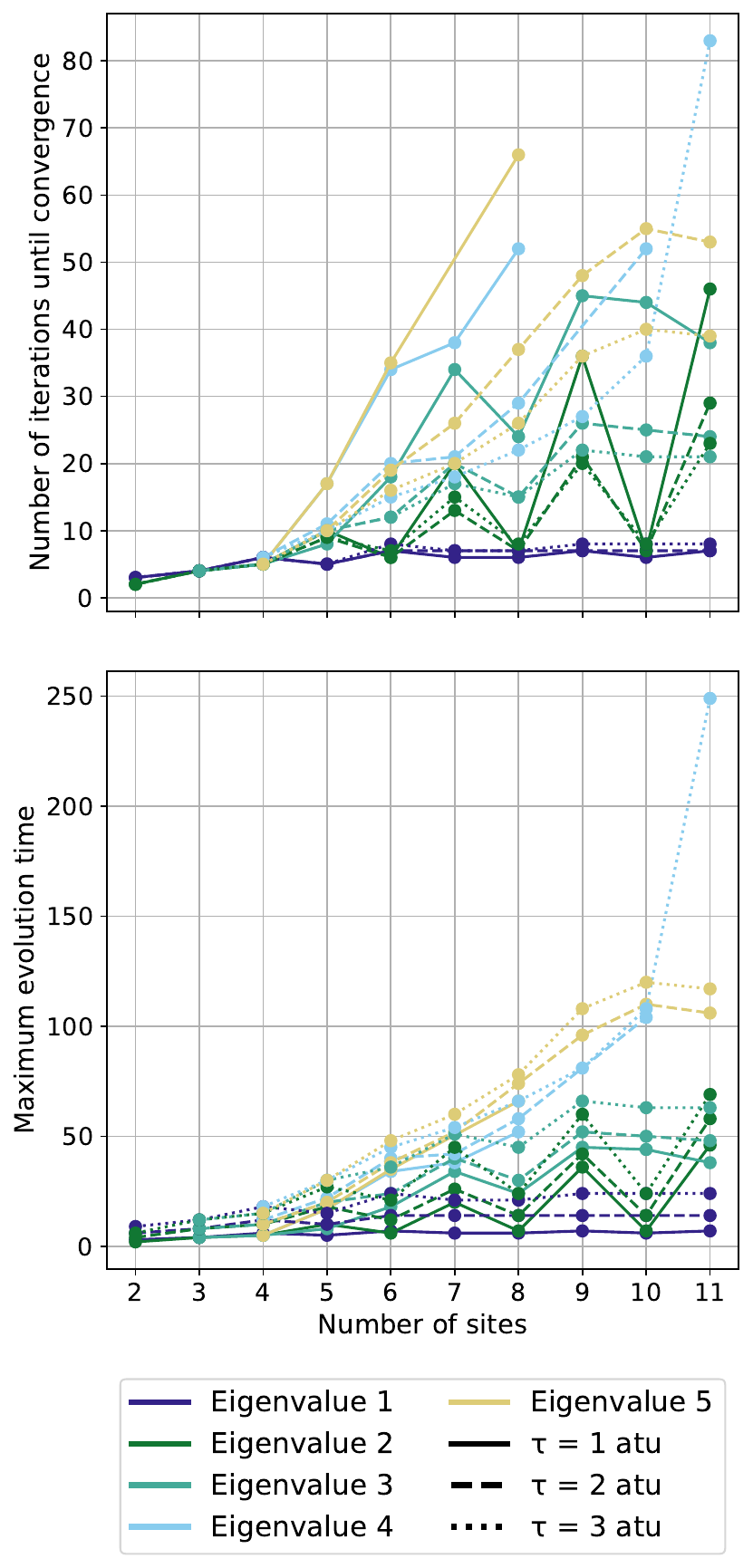}
    \caption{Convergence of the QBKSP algorithm for the Heisenberg model regarding the number of iterations (on the top panel) and the maximum duration of the time evolution required (on the bottom panel) as a function of the problem size. 
    In both panels, we show the convergence of the five lowest (different) eigenvalues with two initial reference states with a fixed overlap of $0.5$ each, for a time evolution duration of one, two, and three atu.
    The maximum number of iterations is set to 200, meaning that if convergence is not achieved within 200 Krylov iterations for larger system sizes, no data point is obtained.}
    \label{fig:heisenberg_convevol}
\end{figure}

\begin{figure*}[htb!]
    \centering
    \includegraphics[width=1\linewidth]{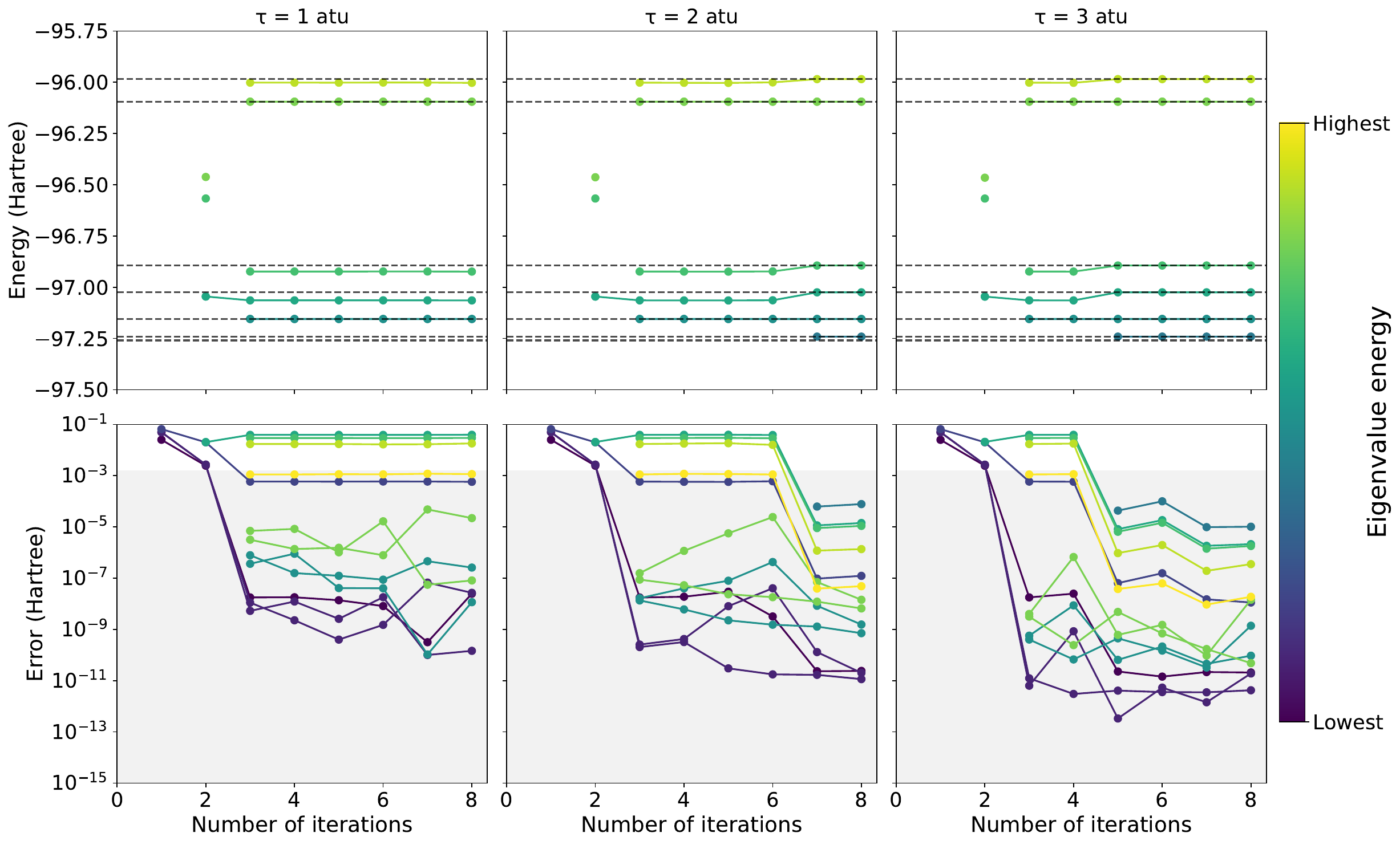}
    \caption{QBKSP algorithm applied to hydrogen fluoride for different durations of the evolution time and four initial reference states, i.e., the HF state and its three dipole-modified variants.
    The top panel shows the convergence of the interior eigenvalues as a function of the Krylov iteration since these are the most challenging to resolve, whereas the bottom panel displays the absolute error of all eigenvalues.
    The grey region indicates values below chemical accuracy.}
    \label{fig:hfev}
\end{figure*}

Since the time step $\tau$ is a key parameter in the QBKSP algorithm, we study its influence on both the quality of the results and the required resources.
The impact of the time evolution duration is analyzed both for model systems (\autoref{fig:heisenberg_convevol}) and molecular Hamiltonians (\autoref{fig:hfev}).

\autoref{fig:heisenberg_convevol} compares the QBKSP convergence obtained with three different time steps $\tau$ for the five lowest eigenvalues of the Heisenberg model with increasing lattice sizes. 
The convergence is shown both in terms of the number of iterations and the maximum evolution time required. 
It is apparent that for an evolution time of 1 atu, larger systems fail to converge to higher-energy eigenvalues within a maximum of 200 Krylov iterations. 
Therefore, to retrieve higher eigenvalues using a fixed number of initial states, one should opt for longer evolution times.
This result is consistent with the phase cancellation conditions presented in Ref. \cite{klymko_real-time_2022}, which are highly likely to be satisfied when longer evolution times are used.
This observation can be intuitively interpreted as a consequence of (auto)correlation functions effectively encoding the full spectral information in the limit of long evolution times, whereas short time evolutions result in only very limited spectral resolution.

In \autoref{fig:hfev}, we illustrate the impact of three different durations of the evolution time ($\tau=1, 2,$ or $3$ atu) on hydrogen fluoride.
For this analysis, we employ four initial reference states, namely the HF state and its three dipole-excited variants ($\hat{\mu}_x$, $\hat{\mu}_y$, and $\hat{\mu}_z$).
We observe that longer evolution times lead to faster convergence, requiring fewer iterations and, consequently, fewer distinct quantum circuits to be executed.

\subsection{Finite sampling effects}
\label{subsec:precision}

\begin{figure*}[htb!]
    \centering
    \includegraphics[width=1\linewidth]{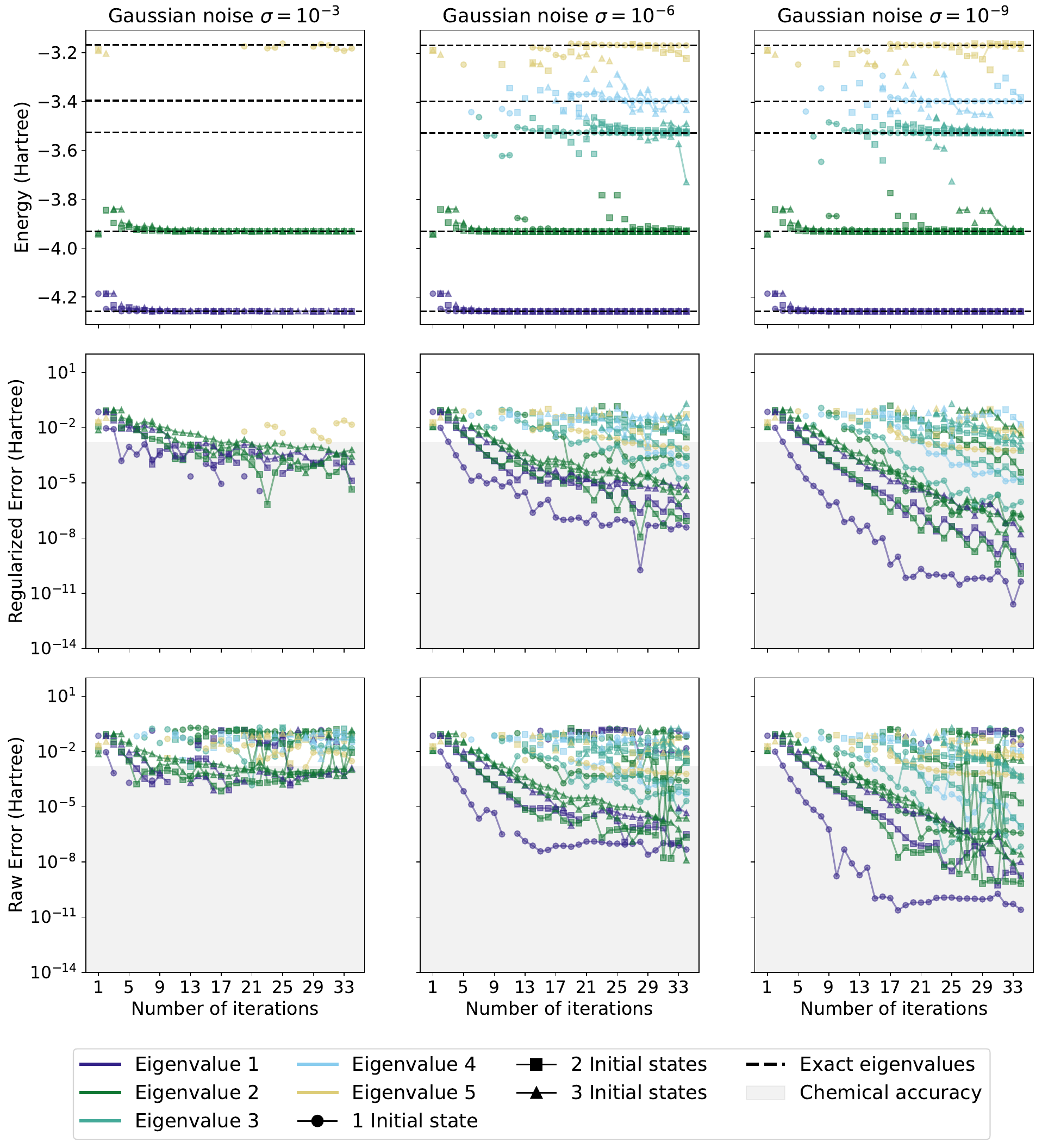}
    \caption{QBKSP algorithm applied to the 10-site Heisenberg model for several magnitudes of Gaussian sampling noise for different numbers of initial reference states with a fixed overlap of $0.5$. 
    The threshold parameter $\epsilon$ of the $S$ matrix regularisation is set to $100\sigma$ to prune numerical errors, with $\sigma$ being the standard deviation of the Gaussian sampling noise. 
    The top panel shows the convergence of the five lowest eigenvalues as a function of Krylov iterations, the middle panel displays the absolute error of the regularized eigenvalues, and the bottom panel shows the absolute error of these eigenvalues before regularization.
    Note that in the leftmost error panels, there are no green circles within the chemical accuracy region, indicating that in noisy regimes adding reference states allows to retrieve higher energy states.}
    \label{fig:Heisenbergpre}
\end{figure*}

Even in an ideal scenario where quantum computers are completely noise-free or perfectly error-corrected, the finite number of shots to measure the different circuits inherently limits the accuracy of the obtained expectation values.
Therefore, as an intermediate step between ideal numerical simulations and quantum circuit simulations, we perform simulations with Gaussian sampling noise to understand the impact of limited quantum circuit shot budgets on the results.
In \autoref{fig:Heisenbergpre}, we show the convergence of the QBKSP algorithm for the 10-site Heisenberg model for different numbers of initial states for three different magnitudes of sampling noise.

This example demonstrates that as the precision of our matrix elements is reduced, both the accuracy and the number of eigenvalues identified decrease.
However, by using several initial reference states, i.e., at least two reference states for the Heisenberg system, it is possible to resolve eigenvalues in the presence of sampling noise with errors below chemical accuracy, which would otherwise be inaccessible with fewer reference states.
Furthermore, it is evident that employing an adequate regularization improves the convergence of the algorithm.

\subsection{Quantum simulations}
\label{subsec:quantumsim}

\begin{figure}[htb!]
    \centering
    \includegraphics[width=1\linewidth]{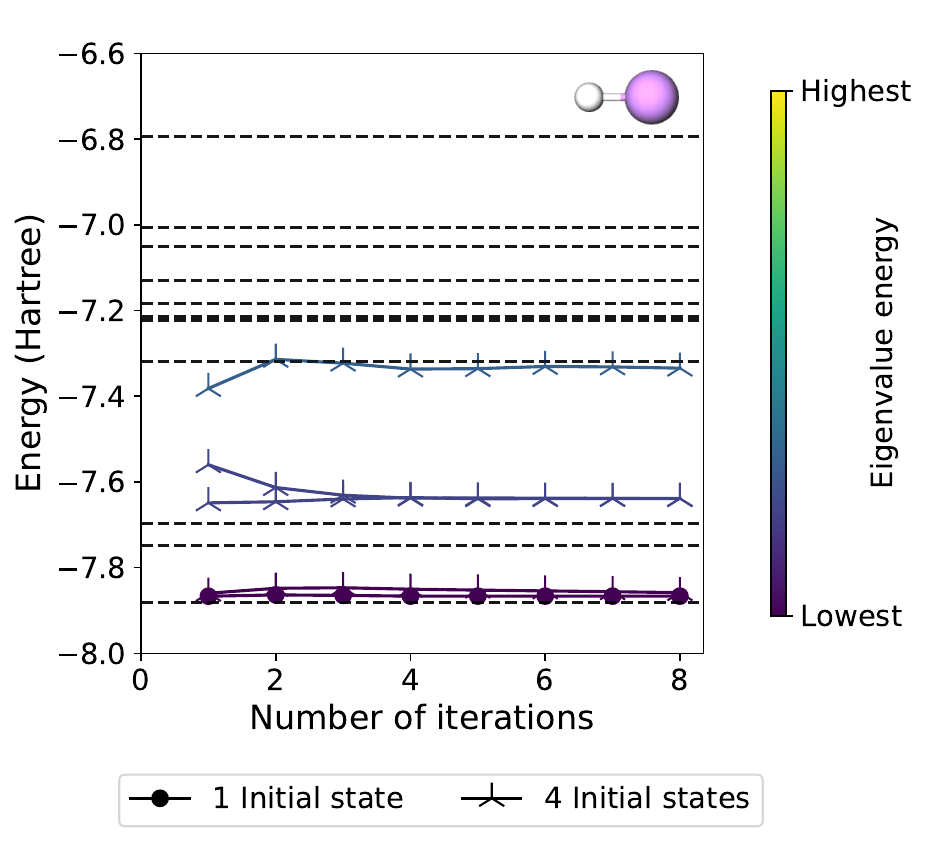}
    \caption{ 
    QBKSP algorithm executed on the Qiskit \textit{Qasm Simulator} applied to the LiH molecule for $10^5$ shots for both one and four references, with a Trotter step size of $0.07$ atu.
    The threshold parameter of the $S$ matrix is set to $0.1$ to reduce numerical instabilities. 
    }
    \label{fig:LiHshots}
\end{figure}

In addition to numerical simulations, we also perform quantum simulations of the LiH molecule for both one reference state (the HF state) and four reference states (the HF and its three dipole-excited variants) using Qiskit \cite{qiskit2024}.
In \autoref{fig:LiHshots} the convergence of the eigenenergies is examined. 
In these simulations, the matrix threshold parameter $S$ is set to $0.1$, the Trotter step size is $0.07$, the number of shots is set to $10^5$, and the total time evolution duration is fixed to $\tau=1$ atu due to computational constraints.
The choice of Trotter step size within the selected parameter range has little effect on the results, as shown in Appendix \ref{app:trotter}.
This is expected, as the noise introduced through the finite sampling of the matrix elements dominates the Trotter error in this setup.

In \autoref{fig:LiHshots}, we have sampling errors of around $\frac{1}{\sqrt{10^5}}\simeq 0.003$.
Thus, the accuracy obtainable by QBKSP is inherently limited by the precision with which the matrix elements are sampled and the correspondingly high cutoff threshold of $0.1$ applied to regularize the eigenvalue problem.
When comparing the single-reference calculation to the QBKSP one with four reference states, it is apparent that using four references leads to the retrieval of a larger number of eigenvalues.

\section{Discussion and conclusions}
\label{sec:discu}

In this work, we present and analyse the QBKSP algorithm, a multireference iterative quantum eigensolver which unifies some of the developments made  in Refs.~\cite{parrish_quantum_2019,stair_multireference_2020,klymko_real-time_2022} and \cite{cortes_quantum_2022}. 
It requires a relatively small number of compact quantum circuits compared to variational approaches such as SSVQE \cite{nakanishi_subspace-search_2019} as convergence is usually achieved within very few Krylov iterations, and it only needs a single ancilla qubit.
The number of circuits required by the QBKSP scales linearly with the number of Krylov iterations, a significant reduction compared to the quadratic scaling of the multireference Krylov method introduced in Ref.~\cite{stair_multireference_2020}.
For the systems investigated in this work, i.e., the Heisenberg model and the lithium hydride and hydrogen fluoride molecules, we observe that the number of Krylov iterations required for converging low-lying eigenstates scales strongly sub-exponentially with the system size for any of the tested numbers of reference states.
When compared to the single-reference approach at the same computational cost, this multireference formulation improves the accuracy of the computed energies, especially for higher-lying eigenstates and in more realistic noisy regimes.
Additionally, QBKSP successfully resolves degenerate states and eigenvalues that are difficult or impossible to capture simultaneously with a single reference.
Although QBKSP requires a maximal evaluation of $B\left(B+1\right)$ different quantum circuits instead of just $2$ for the single reference case in each Krylov iteration, with $B$ being the number of reference states, the execution of these quantum circuits is trivially parallelizable.
Moreover, a small number of references is typically sufficient to realize the advantages of the multireference approach.
As future work, more sophisticated reference selection techniques, such as an extension of the scheme proposed in Ref.~\cite{stair_multireference_2020} for excited states, could be applied in order to further reduce the number of Krylov basis states required to span the subspace of interest.

From the analysis regarding the initial states, we observe that increasing the number of initial reference states improves the accuracy of higher-energy eigenvalue estimates.
For the molecular systems studied in this work, the use of four reference states enables the retrieval of several eigenenergies and respective degeneracies with high resolution within a few iterations.
In particular, we retrieve 10 out of 11 eigenstates for lithium hydride with the correct multiplicity of degeneracies with QBKSP within a few Krylov iterations, compared to only 4 states when using the HF state as single reference at the same computational cost.
The use of four initial reference states is also advantageous when compared to a single-reference approach corresponding to a uniform superposition of the four individual references, allowing not only finding degenerate states inaccessible with just one reference but also requiring less distinct quantum circuits to retrieve the same number of eigenvalues.
Thus, the use of multiple reference states also facilitates the retrieval of eigenvalues that would otherwise be inaccessible, for instance, due to symmetry restrictions, when relying on a single reference state.
We successfully calculate the lowest and highest eigenenergies of diatomic molecules at different interatomic distances, demonstrating the robustness of the method across varying correlation regimes.
Concerning the effect of the evolution time, as expected, a general trend is observed that under perfect conditions longer evolution times improve both the convergence and the final energy accuracy.
However, implementing longer execution times may be resource-intensive on real quantum devices, which may lead to choosing shorter evolution times and, e.g., improving the accuracy by using more initial reference states.

Realistic precision constraints, such as those arising from sampling a finite number of shots on quantum devices, can significantly impact algorithmic performance.
By performing simulations with Gaussian sampling noise,  we find that restricting numerical precision reduces both the number of recoverable eigenvalues and their accuracy, as expected. 
Nonetheless, by using a block of initial reference states this problem is partially mitigated since using more references allows us to find eigenvalues with errors below the chemical accuracy that otherwise were unreachable.
In the quantum circuit simulations, we qualitatively resolve the ground state and observe that using four reference states improves the convergence and seems to enable the retrieval of excited states. 
We observe that the choice of the truncation threshold for the overlap matrix plays a pivotal role, and quantum simulations, which include both Trotter and sampling errors, require higher threshold values. 
In future work, a similar analysis to the one presented in Refs.~\cite{lee_sampling_2024,lee_efficient_2025} could be performed to determine not only the optimal truncation threshold and number of shots, but also the ideal Trotter step size and evolution time for the QBKSP algorithm. 
Moreover, an extension of the error analysis presented in Ref.~\cite{Kirby_2024} could be used to assess the nonasymptotic error bounds of QBKSP for both ground and excited states.

Consistent with the findings in Ref.~\cite{baker_block_2024}, we observe that when the calculation is already converged for several iterations for model Hamiltonians, the subspace diagonalization can result in multiplicities of degenerate eigenenergies that are higher than the number of degenerate states actually present in the system. 
This artifact can be mitigated by evaluating the energy and multiplicity of each eigenvalue immediately after its convergence, or alternatively by increasing the threshold parameter of the overlap matrix regularization, albeit at the cost of limiting the resulting accuracy. 
In our studies, we observed that using an energy threshold of $0.1$mHa between two subsequent iterations as a convergence criterion reliably yields the correct energy values and respective multiplicities.
Moreover, the corresponding eigenstates exhibit minimal overlaps, whereas the spurious states emerging several iterations after convergence have large overlaps with other eigenstates in the spanned subspace.
Consequently, the overlaps of the degenerate states can serve as straightforward measures to distinguish genuine solutions from spurious states. 
Note that, as a quantum subspace method, the QBKSP algorithm naturally provides access to overlaps between different eigenstates, i.e., without the need for any further quantum computations.

While the primary advantage of the QBKSP algorithm for model Hamiltonians lies in the retrieval of degenerate states, the molecular test cases show additional benefits.
Compared to the single-reference case, the QBKSP algorithm requires fewer quantum circuit evaluations to converge the target eigenvalues within chemical accuracy, and it enables the retrieval of eigenstates that remain inaccessible when using only one reference state.

Another point to highlight is that the QBKSP, as usual for Krylov subspace projector methods, starts by converging to the most extreme eigenvalues, i.e., the ground state and the most excited state. 
This feature enables not only the extraction of low-energy spectra but also the determination of the full spectral range of the Hamiltonian, which can be useful, for instance, to obtain a better estimate of the operator norm for future quantum calculations.

\section*{Availability of data and code}
The data presented in this work was generated using the code that the authors made available at \url{https://github.com/NQCP/NQCP-AA-pub-QBKSP\_eigensolver} \cite{OliveiraGlaser2025} and is available upon reasonable request.

\begin{acknowledgments}
This work is supported by the Novo Nordisk Foundation, Grant number NNF22SA0081175, NNF Quantum Computing Programme.
The authors thank Marcel Fabian for the valuable discussions, and Matthew Teynor and Andreas Bock Michelsen for their helpful comments on the manuscript. M. G. J. O. also thanks Carl Gustav Henning Hansen, Francesco Ferrarin, and Daria Gusew for their feedback on sections of the manuscript.
\end{acknowledgments}



\begin{thebibliography}{56}%
\makeatletter
\providecommand \@ifxundefined [1]{%
 \@ifx{#1\undefined}
}%
\providecommand \@ifnum [1]{%
 \ifnum #1\expandafter \@firstoftwo
 \else \expandafter \@secondoftwo
 \fi
}%
\providecommand \@ifx [1]{%
 \ifx #1\expandafter \@firstoftwo
 \else \expandafter \@secondoftwo
 \fi
}%
\providecommand \natexlab [1]{#1}%
\providecommand \enquote  [1]{``#1''}%
\providecommand \bibnamefont  [1]{#1}%
\providecommand \bibfnamefont [1]{#1}%
\providecommand \citenamefont [1]{#1}%
\providecommand \href@noop [0]{\@secondoftwo}%
\providecommand \href [0]{\begingroup \@sanitize@url \@href}%
\providecommand \@href[1]{\@@startlink{#1}\@@href}%
\providecommand \@@href[1]{\endgroup#1\@@endlink}%
\providecommand \@sanitize@url [0]{\catcode `\\12\catcode `\$12\catcode `\&12\catcode `\#12\catcode `\^12\catcode `\_12\catcode `\%12\relax}%
\providecommand \@@startlink[1]{}%
\providecommand \@@endlink[0]{}%
\providecommand \url  [0]{\begingroup\@sanitize@url \@url }%
\providecommand \@url [1]{\endgroup\@href {#1}{\urlprefix }}%
\providecommand \urlprefix  [0]{URL }%
\providecommand \Eprint [0]{\href }%
\providecommand \doibase [0]{https://doi.org/}%
\providecommand \selectlanguage [0]{\@gobble}%
\providecommand \bibinfo  [0]{\@secondoftwo}%
\providecommand \bibfield  [0]{\@secondoftwo}%
\providecommand \translation [1]{[#1]}%
\providecommand \BibitemOpen [0]{}%
\providecommand \bibitemStop [0]{}%
\providecommand \bibitemNoStop [0]{.\EOS\space}%
\providecommand \EOS [0]{\spacefactor3000\relax}%
\providecommand \BibitemShut  [1]{\csname bibitem#1\endcsname}%
\let\auto@bib@innerbib\@empty
\bibitem [{\citenamefont {Thouless}(1972)}]{thouless_quantum_1972}%
  \BibitemOpen
  \bibfield  {author} {\bibinfo {author} {\bibfnamefont {D.~J.}\ \bibnamefont {Thouless}},\ } {} {\emph {\bibinfo {title} {The {Quantum} {Mechanics} of {Many}-body {Systems}}}}\ (\bibinfo  {publisher} {Academic Press,  New York},\ \bibinfo {year} {1972})\BibitemShut {NoStop}%
\bibitem [{\citenamefont {Feynman}(1982)}]{feynman_simulating_1982}%
  \BibitemOpen
  \bibfield  {author} {\bibinfo {author} {\bibfnamefont {R.~P.}\ \bibnamefont {Feynman}},\ }\bibfield  {title} {\bibinfo {title} {Simulating physics with computers},\ }\href {https://doi.org/10.1007/BF02650179} {\bibfield  {journal} {\bibinfo  {journal} {Int. J. Theor. Phys.}\ }\textbf {\bibinfo {volume} {21}},\ \bibinfo {pages} {467} (\bibinfo {year} {1982})}\BibitemShut {NoStop}%
\bibitem [{\citenamefont {Francis}(1961)}]{francis_qr_1961}%
  \BibitemOpen
  \bibfield  {author} {\bibinfo {author} {\bibfnamefont {J.~G.~F.}\ \bibnamefont {Francis}},\ }\bibfield  {title} {\bibinfo {title} {The {QR} {Transformation} {A} {Unitary} {Analogue} to the {LR} {Transformation}—{Part} 1},\ }\href {https://doi.org/10.1093/comjnl/4.3.265} {\bibfield  {journal} {\bibinfo  {journal} {Comput. J.}\ }\textbf {\bibinfo {volume} {4}},\ \bibinfo {pages} {265} (\bibinfo {year} {1961})}\BibitemShut {NoStop}%
\bibitem [{\citenamefont {Francis}(1962)}]{francis_qr_1962}%
  \BibitemOpen
  \bibfield  {author} {\bibinfo {author} {\bibfnamefont {J.~G.~F.}\ \bibnamefont {Francis}},\ }\bibfield  {title} {\bibinfo {title} {The {QR} {Transformation}—{Part} 2},\ }\href {https://doi.org/10.1093/comjnl/4.4.332} {\bibfield  {journal} {\bibinfo  {journal} {Comput. J.}\ }\textbf {\bibinfo {volume} {4}},\ \bibinfo {pages} {332} (\bibinfo {year} {1962})}\BibitemShut {NoStop}%
\bibitem [{\citenamefont {Kublanovskaya}(1962)}]{kublanovskaya_algorithms_1962}%
  \BibitemOpen
  \bibfield  {author} {\bibinfo {author} {\bibfnamefont {V.~N.}\ \bibnamefont {Kublanovskaya}},\ }\bibfield  {title} {\bibinfo {title} {On some algorithms for the solution of the complete eigenvalue problem},\ }\href {https://doi.org/10.1016/0041-5553(63)90168-X} {\bibfield  {journal} {\bibinfo  {journal} {USSR Comput. Math. Math. Phys.}\ }\textbf {\bibinfo {volume} {1}},\ \bibinfo {pages} {637} (\bibinfo {year} {1962})}\BibitemShut {NoStop}%
\bibitem [{\citenamefont {Hoffmann}(1989)}]{hoffmann_iterative_1989}%
  \BibitemOpen
  \bibfield  {author} {\bibinfo {author} {\bibfnamefont {W.}~\bibnamefont {Hoffmann}},\ }\bibfield  {title} {\bibinfo {title} {Iterative algorithms for {Gram}-{Schmidt} orthogonalization},\ }\href {https://doi.org/10.1007/BF02241222} {\bibfield  {journal} {\bibinfo  {journal} {Computing}\ }\textbf {\bibinfo {volume} {41}},\ \bibinfo {pages} {335} (\bibinfo {year} {1989})}\BibitemShut {NoStop}%
\bibitem [{\citenamefont {Leissa}(2005)}]{leissa_historical_2005}%
  \BibitemOpen
  \bibfield  {author} {\bibinfo {author} {\bibfnamefont {A.~W.}\ \bibnamefont {Leissa}},\ }\bibfield  {title} {\bibinfo {title} {The historical bases of the {Rayleigh} and {Ritz} methods},\ }\href {https://doi.org/10.1016/j.jsv.2004.12.021} {\bibfield  {journal} {\bibinfo  {journal} {J. Sound Vib.}\ }\textbf {\bibinfo {volume} {287}},\ \bibinfo {pages} {961} (\bibinfo {year} {2005})}\BibitemShut {NoStop}%
\bibitem [{\citenamefont {Bai}\ \emph {et~al.}(2021)\citenamefont {Bai}, \citenamefont {Wu},\ and\ \citenamefont {Muratova}}]{bai_power_2021}%
  \BibitemOpen
  \bibfield  {author} {\bibinfo {author} {\bibfnamefont {Z.-Z.}\ \bibnamefont {Bai}}, \bibinfo {author} {\bibfnamefont {W.-T.}\ \bibnamefont {Wu}},\ and\ \bibinfo {author} {\bibfnamefont {G.~V.}\ \bibnamefont {Muratova}},\ }\bibfield  {title} {\bibinfo {title} {The power method and beyond},\ }\href {https://doi.org/10.1016/j.apnum.2020.03.021} {\bibfield  {journal} {\bibinfo  {journal} {Appl. Numer. Math.}\ }\textbf {\bibinfo {volume} {164}},\ \bibinfo {pages} {29} (\bibinfo {year} {2021})}\BibitemShut {NoStop}%
\bibitem [{\citenamefont {Arnoldi}(1951)}]{arnoldi_principle_1951}%
  \BibitemOpen
  \bibfield  {author} {\bibinfo {author} {\bibfnamefont {W.~E.}\ \bibnamefont {Arnoldi}},\ }\bibfield  {title} {\bibinfo {title} {The principle of minimized iterations in the solution of the matrix eigenvalue problem},\ }\href {https://doi.org/10.1090/qam/42792} {\bibfield  {journal} {\bibinfo  {journal} {Q. Appl. Math.}\ }\textbf {\bibinfo {volume} {9}},\ \bibinfo {pages} {17} (\bibinfo {year} {1951})}\BibitemShut {NoStop}%
\bibitem [{\citenamefont {Lanczos}(1950)}]{lanczos_iteration_1950}%
  \BibitemOpen
  \bibfield  {author} {\bibinfo {author} {\bibfnamefont {C.}~\bibnamefont {Lanczos}},\ }\bibfield  {title} {\bibinfo {title} {An iteration method for the solution of the eigenvalue problem of linear differential and integral operators},\ }\href {https://doi.org/10.6028/jres.045.026} {\bibfield  {journal} {\bibinfo  {journal} {J. Res. Natl. Bur. Stand.}\ }\textbf {\bibinfo {volume} {45}},\ \bibinfo {pages} {255} (\bibinfo {year} {1950})}\BibitemShut {NoStop}%
\bibitem [{\citenamefont {Grosso}\ \emph {et~al.}(1995)\citenamefont {Grosso}, \citenamefont {Martinelli},\ and\ \citenamefont {Pastori~Parravicini}}]{grosso_lanczos-type_1995}%
  \BibitemOpen
  \bibfield  {author} {\bibinfo {author} {\bibfnamefont {G.}~\bibnamefont {Grosso}}, \bibinfo {author} {\bibfnamefont {L.}~\bibnamefont {Martinelli}},\ and\ \bibinfo {author} {\bibfnamefont {G.}~\bibnamefont {Pastori~Parravicini}},\ }\bibfield  {title} {\bibinfo {title} {Lanczos-type algorithm for excited states of very-large-scale quantum systems},\ }\href {https://doi.org/10.1103/physrevb.51.13033} {\bibfield  {journal} {\bibinfo  {journal} {Phys. Rev. B}\ }\textbf {\bibinfo {volume} {51}},\ \bibinfo {pages} {13033} (\bibinfo {year} {1995})}\BibitemShut {NoStop}%
\bibitem [{\citenamefont {Grüning}\ \emph {et~al.}(2011)\citenamefont {Grüning}, \citenamefont {Marini},\ and\ \citenamefont {Gonze}}]{gruning_implementation_2011}%
  \BibitemOpen
  \bibfield  {author} {\bibinfo {author} {\bibfnamefont {M.}~\bibnamefont {Grüning}}, \bibinfo {author} {\bibfnamefont {A.}~\bibnamefont {Marini}},\ and\ \bibinfo {author} {\bibfnamefont {X.}~\bibnamefont {Gonze}},\ }\bibfield  {title} {\bibinfo {title} {Implementation and testing of {Lanczos}-based algorithms for {Random}-{Phase} {Approximation} eigenproblems},\ }\href {https://doi.org/10.1016/j.commatsci.2011.02.021} {\bibfield  {journal} {\bibinfo  {journal} {Comput. Mater. Sci.}\ }\textbf {\bibinfo {volume} {50}},\ \bibinfo {pages} {2148} (\bibinfo {year} {2011})}\BibitemShut {NoStop}%
\bibitem [{\citenamefont {Kitaev}(1995)}]{kitaev_quantum_1995}%
  \BibitemOpen
  \bibfield  {author} {\bibinfo {author} {\bibfnamefont {A.}~\bibnamefont {Kitaev}},\ }\bibfield  {title} {\bibinfo {title} {Quantum measurements and the {Abelian} {Stabilizer} {Problem}},\ }\href {https://www.semanticscholar.org/paper/Quantum-measurements-and-the-Abelian-Stabilizer-Kitaev/e218de049ab533e7d54f336cadac942effddf139} {\bibfield  {journal} {\bibinfo  {journal} {ECCC}\ }\textbf {\bibinfo {volume} {TR96}} (\bibinfo {year} {1995})}\BibitemShut {NoStop}%
\bibitem [{\citenamefont {O'Loan}(2009)}]{oloan_iterative_2009}%
  \BibitemOpen
  \bibfield  {author} {\bibinfo {author} {\bibfnamefont {C.~J.}\ \bibnamefont {O'Loan}},\ }\bibfield  {title} {\bibinfo {title} {Iterative phase estimation},\ }\href {https://doi.org/10.1088/1751-8113/43/1/015301} {\bibfield  {journal} {\bibinfo  {journal} {J. Phys. A: Math. Theor.}\ }\textbf {\bibinfo {volume} {43}},\ \bibinfo {pages} {015301} (\bibinfo {year} {2010})}\BibitemShut {NoStop}%
\bibitem [{\citenamefont {Peruzzo}\ \emph {et~al.}(2014)\citenamefont {Peruzzo}, \citenamefont {McClean}, \citenamefont {Shadbolt}, \citenamefont {Yung}, \citenamefont {Zhou}, \citenamefont {Love}, \citenamefont {Aspuru-Guzik},\ and\ \citenamefont {O’Brien}}]{peruzzo_variational_2014}%
  \BibitemOpen
  \bibfield  {author} {\bibinfo {author} {\bibfnamefont {A.}~\bibnamefont {Peruzzo}}, \bibinfo {author} {\bibfnamefont {J.}~\bibnamefont {McClean}}, \bibinfo {author} {\bibfnamefont {P.}~\bibnamefont {Shadbolt}}, \bibinfo {author} {\bibfnamefont {M.-H.}\ \bibnamefont {Yung}}, \bibinfo {author} {\bibfnamefont {X.-Q.}\ \bibnamefont {Zhou}}, \bibinfo {author} {\bibfnamefont {P.~J.}\ \bibnamefont {Love}}, \bibinfo {author} {\bibfnamefont {A.}~\bibnamefont {Aspuru-Guzik}},\ and\ \bibinfo {author} {\bibfnamefont {J.~L.}\ \bibnamefont {O’Brien}},\ }\bibfield  {title} {\bibinfo {title} {A variational eigenvalue solver on a photonic quantum processor},\ }\href {https://doi.org/10.1038/ncomms5213} {\bibfield  {journal} {\bibinfo  {journal} {Nat. Commun.}\ }\textbf {\bibinfo {volume} {5}},\ \bibinfo {pages} {4213} (\bibinfo {year} {2014})}\BibitemShut {NoStop}%
\bibitem [{\citenamefont {Tilly}\ \emph {et~al.}(2022)\citenamefont {Tilly}, \citenamefont {Chen}, \citenamefont {Cao}, \citenamefont {Picozzi}, \citenamefont {Setia}, \citenamefont {Li}, \citenamefont {Grant}, \citenamefont {Wossnig}, \citenamefont {Rungger}, \citenamefont {Booth},\ and\ \citenamefont {Tennyson}}]{tilly_variational_2022}%
  \BibitemOpen
  \bibfield  {author} {\bibinfo {author} {\bibfnamefont {J.}~\bibnamefont {Tilly}}, \bibinfo {author} {\bibfnamefont {H.}~\bibnamefont {Chen}}, \bibinfo {author} {\bibfnamefont {S.}~\bibnamefont {Cao}}, \bibinfo {author} {\bibfnamefont {D.}~\bibnamefont {Picozzi}}, \bibinfo {author} {\bibfnamefont {K.}~\bibnamefont {Setia}}, \bibinfo {author} {\bibfnamefont {Y.}~\bibnamefont {Li}}, \bibinfo {author} {\bibfnamefont {E.}~\bibnamefont {Grant}}, \bibinfo {author} {\bibfnamefont {L.}~\bibnamefont {Wossnig}}, \bibinfo {author} {\bibfnamefont {I.}~\bibnamefont {Rungger}}, \bibinfo {author} {\bibfnamefont {G.~H.}\ \bibnamefont {Booth}},\ and\ \bibinfo {author} {\bibfnamefont {J.}~\bibnamefont {Tennyson}},\ }\bibfield  {title} {\bibinfo {title} {The {Variational} {Quantum} {Eigensolver}: a review of methods and best practices},\ }\href {https://doi.org/10.1016/j.physrep.2022.08.003} {\bibfield  {journal} {\bibinfo  {journal} {Phys. Rep.}\ }\textbf {\bibinfo {volume} {986}},\ \bibinfo {pages} {1} (\bibinfo {year}
  {2022})}\BibitemShut {NoStop}%
\bibitem [{\citenamefont {Nakanishi}\ \emph {et~al.}(2019)\citenamefont {Nakanishi}, \citenamefont {Mitarai},\ and\ \citenamefont {Fujii}}]{nakanishi_subspace-search_2019}%
  \BibitemOpen
  \bibfield  {author} {\bibinfo {author} {\bibfnamefont {K.~M.}\ \bibnamefont {Nakanishi}}, \bibinfo {author} {\bibfnamefont {K.}~\bibnamefont {Mitarai}},\ and\ \bibinfo {author} {\bibfnamefont {K.}~\bibnamefont {Fujii}},\ }\bibfield  {title} {\bibinfo {title} {Subspace-search variational quantum eigensolver for excited states},\ }\href {https://doi.org/10.1103/PhysRevResearch.1.033062} {\bibfield  {journal} {\bibinfo  {journal} {Phys. Rev. Res.}\ }\textbf {\bibinfo {volume} {1}},\ \bibinfo {pages} {033062} (\bibinfo {year} {2019})}\BibitemShut {NoStop}%
\bibitem [{\citenamefont {Ding}\ \emph {et~al.}(2024)\citenamefont {Ding}, \citenamefont {Li}, \citenamefont {Lin}, \citenamefont {Ni}, \citenamefont {Ying},\ and\ \citenamefont {Zhang}}]{ding_quantum_2024}%
  \BibitemOpen
  \bibfield  {author} {\bibinfo {author} {\bibfnamefont {Z.}~\bibnamefont {Ding}}, \bibinfo {author} {\bibfnamefont {H.}~\bibnamefont {Li}}, \bibinfo {author} {\bibfnamefont {L.}~\bibnamefont {Lin}}, \bibinfo {author} {\bibfnamefont {H.}~\bibnamefont {Ni}}, \bibinfo {author} {\bibfnamefont {L.}~\bibnamefont {Ying}},\ and\ \bibinfo {author} {\bibfnamefont {R.}~\bibnamefont {Zhang}},\ }\bibfield  {title} {\bibinfo {title} {Quantum {Multiple} {Eigenvalue} {Gaussian} filtered {Search}: an efficient and versatile quantum phase estimation method},\ }\href {https://doi.org/10.22331/q-2024-10-02-1487} {\bibfield  {journal} {\bibinfo  {journal} {Quantum}\ }\textbf {\bibinfo {volume} {8}},\ \bibinfo {pages} {1487} (\bibinfo {year} {2024})}\BibitemShut {NoStop}%
\bibitem [{\citenamefont {Ding}\ and\ \citenamefont {Lin}(2023)}]{ding_simultaneous_2023}%
  \BibitemOpen
  \bibfield  {author} {\bibinfo {author} {\bibfnamefont {Z.}~\bibnamefont {Ding}}\ and\ \bibinfo {author} {\bibfnamefont {L.}~\bibnamefont {Lin}},\ }\bibfield  {title} {\bibinfo {title} {Simultaneous estimation of multiple eigenvalues with short-depth quantum circuit on early fault-tolerant quantum computers},\ }\href {https://doi.org/10.22331/q-2023-10-11-1136} {\bibfield  {journal} {\bibinfo  {journal} {Quantum}\ }\textbf {\bibinfo {volume} {7}},\ \bibinfo {pages} {1136} (\bibinfo {year} {2023})}\BibitemShut {NoStop}%
\bibitem [{\citenamefont {Parrish}\ and\ \citenamefont {McMahon}(2019)}]{parrish_quantum_2019}%
  \BibitemOpen
  \bibfield  {author} {\bibinfo {author} {\bibfnamefont {R.~M.}\ \bibnamefont {Parrish}}\ and\ \bibinfo {author} {\bibfnamefont {P.~L.}\ \bibnamefont {McMahon}},\ }\href {https://doi.org/10.48550/arXiv.1909.08925} {\bibinfo {title} {Quantum {Filter} {Diagonalization}: {Quantum} {Eigendecomposition} without {Full} {Quantum} {Phase} {Estimation}}} (\bibinfo {year} {2019}),\ \bibinfo {note} {10.48550/arXiv.1909.08925}\BibitemShut {NoStop}%
\bibitem [{\citenamefont {Stair}\ \emph {et~al.}(2020)\citenamefont {Stair}, \citenamefont {Huang},\ and\ \citenamefont {Evangelista}}]{stair_multireference_2020}%
  \BibitemOpen
  \bibfield  {author} {\bibinfo {author} {\bibfnamefont {N.~H.}\ \bibnamefont {Stair}}, \bibinfo {author} {\bibfnamefont {R.}~\bibnamefont {Huang}},\ and\ \bibinfo {author} {\bibfnamefont {F.~A.}\ \bibnamefont {Evangelista}},\ }\bibfield  {title} {\bibinfo {title} {A {Multireference} {Quantum} {Krylov} {Algorithm} for {Strongly} {Correlated} {Electrons}},\ }\href {https://doi.org/10.1021/acs.jctc.9b01125} {\bibfield  {journal} {\bibinfo  {journal} {J. Chem. Theory Comput.}\ }\textbf {\bibinfo {volume} {16}},\ \bibinfo {pages} {2236} (\bibinfo {year} {2020})}\BibitemShut {NoStop}%
\bibitem [{\citenamefont {Klymko}\ \emph {et~al.}(2022)\citenamefont {Klymko}, \citenamefont {Mejuto-Zaera}, \citenamefont {Cotton}, \citenamefont {Wudarski}, \citenamefont {Urbanek}, \citenamefont {Hait}, \citenamefont {Head-Gordon}, \citenamefont {Whaley}, \citenamefont {Moussa}, \citenamefont {Wiebe}, \citenamefont {de~Jong},\ and\ \citenamefont {Tubman}}]{klymko_real-time_2022}%
  \BibitemOpen
  \bibfield  {author} {\bibinfo {author} {\bibfnamefont {K.}~\bibnamefont {Klymko}}, \bibinfo {author} {\bibfnamefont {C.}~\bibnamefont {Mejuto-Zaera}}, \bibinfo {author} {\bibfnamefont {S.~J.}\ \bibnamefont {Cotton}}, \bibinfo {author} {\bibfnamefont {F.}~\bibnamefont {Wudarski}}, \bibinfo {author} {\bibfnamefont {M.}~\bibnamefont {Urbanek}}, \bibinfo {author} {\bibfnamefont {D.}~\bibnamefont {Hait}}, \bibinfo {author} {\bibfnamefont {M.}~\bibnamefont {Head-Gordon}}, \bibinfo {author} {\bibfnamefont {K.~B.}\ \bibnamefont {Whaley}}, \bibinfo {author} {\bibfnamefont {J.}~\bibnamefont {Moussa}}, \bibinfo {author} {\bibfnamefont {N.}~\bibnamefont {Wiebe}}, \bibinfo {author} {\bibfnamefont {W.~A.}\ \bibnamefont {de~Jong}},\ and\ \bibinfo {author} {\bibfnamefont {N.~M.}\ \bibnamefont {Tubman}},\ }\bibfield  {title} {\bibinfo {title} {Real-{Time} {Evolution} for {Ultracompact} {Hamiltonian} {Eigenstates} on {Quantum} {Hardware}},\ }\href {https://doi.org/10.1103/PRXQuantum.3.020323} {\bibfield  {journal} {\bibinfo
  {journal} {PRX Quantum}\ }\textbf {\bibinfo {volume} {3}},\ \bibinfo {pages} {020323} (\bibinfo {year} {2022})}\BibitemShut {NoStop}%
\bibitem [{\citenamefont {Cortes}\ and\ \citenamefont {Gray}(2022)}]{cortes_quantum_2022}%
  \BibitemOpen
  \bibfield  {author} {\bibinfo {author} {\bibfnamefont {C.~L.}\ \bibnamefont {Cortes}}\ and\ \bibinfo {author} {\bibfnamefont {S.~K.}\ \bibnamefont {Gray}},\ }\bibfield  {title} {\bibinfo {title} {Quantum {Krylov} subspace algorithms for ground and excited state energy estimation},\ }\href {https://doi.org/10.1103/PhysRevA.105.022417} {\bibfield  {journal} {\bibinfo  {journal} {Phys. Rev. A}\ }\textbf {\bibinfo {volume} {105}},\ \bibinfo {pages} {022417} (\bibinfo {year} {2022})}\BibitemShut {NoStop}%
\bibitem [{\citenamefont {Baker}(2024)}]{baker_block_2024}%
  \BibitemOpen
  \bibfield  {author} {\bibinfo {author} {\bibfnamefont {T.~E.}\ \bibnamefont {Baker}},\ }\bibfield  {title} {\bibinfo {title} {Block {Lanczos} method for excited states on a quantum computer},\ }\href {https://doi.org/10.1103/PhysRevA.110.012420} {\bibfield  {journal} {\bibinfo  {journal} {Phys. Rev. A}\ }\textbf {\bibinfo {volume} {110}},\ \bibinfo {pages} {012420} (\bibinfo {year} {2024})}\BibitemShut {NoStop}%
\bibitem [{\citenamefont {Oumarou}\ \emph {et~al.}(2025)\citenamefont {Oumarou}, \citenamefont {Ollitrault}, \citenamefont {Cortes}, \citenamefont {Scheurer}, \citenamefont {Parrish},\ and\ \citenamefont {Gogolin}}]{oumarou_molecular_2025}%
  \BibitemOpen
  \bibfield  {author} {\bibinfo {author} {\bibfnamefont {O.}~\bibnamefont {Oumarou}}, \bibinfo {author} {\bibfnamefont {P.~J.}\ \bibnamefont {Ollitrault}}, \bibinfo {author} {\bibfnamefont {C.~L.}\ \bibnamefont {Cortes}}, \bibinfo {author} {\bibfnamefont {M.}~\bibnamefont {Scheurer}}, \bibinfo {author} {\bibfnamefont {R.~M.}\ \bibnamefont {Parrish}},\ and\ \bibinfo {author} {\bibfnamefont {C.}~\bibnamefont {Gogolin}},\ }\bibfield  {title} {\bibinfo {title} {Molecular {Properties} from {Quantum} {Krylov} {Subspace} {Diagonalization}},\ }\href {https://doi.org/10.1021/acs.jctc.5c00194} {\bibfield  {journal} {\bibinfo  {journal} {J. Chem. Theory Comput.}\ }\textbf {\bibinfo {volume} {21}},\ \bibinfo {pages} {4543} (\bibinfo {year} {2025})}\BibitemShut {NoStop}%
\bibitem [{\citenamefont {Baker}(2021)}]{baker_lanczos_2021}%
  \BibitemOpen
  \bibfield  {author} {\bibinfo {author} {\bibfnamefont {T.~E.}\ \bibnamefont {Baker}},\ }\bibfield  {title} {\bibinfo {title} {Lanczos recursion on a quantum computer for the {Green}'s function and ground state},\ }\href {https://doi.org/10.1103/PhysRevA.103.032404} {\bibfield  {journal} {\bibinfo  {journal} {Phys. Rev. A}\ }\textbf {\bibinfo {volume} {103}},\ \bibinfo {pages} {032404} (\bibinfo {year} {2021})}\BibitemShut {NoStop}%
\bibitem [{\citenamefont {Motta}\ \emph {et~al.}(2020)\citenamefont {Motta}, \citenamefont {Sun}, \citenamefont {Tan}, \citenamefont {O’Rourke}, \citenamefont {Ye}, \citenamefont {Minnich}, \citenamefont {Brandão},\ and\ \citenamefont {Chan}}]{motta_determining_2020}%
  \BibitemOpen
  \bibfield  {author} {\bibinfo {author} {\bibfnamefont {M.}~\bibnamefont {Motta}}, \bibinfo {author} {\bibfnamefont {C.}~\bibnamefont {Sun}}, \bibinfo {author} {\bibfnamefont {A.~T.~K.}\ \bibnamefont {Tan}}, \bibinfo {author} {\bibfnamefont {M.~J.}\ \bibnamefont {O’Rourke}}, \bibinfo {author} {\bibfnamefont {E.}~\bibnamefont {Ye}}, \bibinfo {author} {\bibfnamefont {A.~J.}\ \bibnamefont {Minnich}}, \bibinfo {author} {\bibfnamefont {F.~G. S.~L.}\ \bibnamefont {Brandão}},\ and\ \bibinfo {author} {\bibfnamefont {G.~K.-L.}\ \bibnamefont {Chan}},\ }\bibfield  {title} {\bibinfo {title} {Determining eigenstates and thermal states on a quantum computer using quantum imaginary time evolution},\ }\href {https://doi.org/10.1038/s41567-019-0704-4} {\bibfield  {journal} {\bibinfo  {journal} {Nat. Phys.}\ }\textbf {\bibinfo {volume} {16}},\ \bibinfo {pages} {205} (\bibinfo {year} {2020})}\BibitemShut {NoStop}%
\bibitem [{\citenamefont {Yeter-Aydeniz}\ \emph {et~al.}(2020)\citenamefont {Yeter-Aydeniz}, \citenamefont {Pooser},\ and\ \citenamefont {Siopsis}}]{yeter-aydeniz_practical_2020}%
  \BibitemOpen
  \bibfield  {author} {\bibinfo {author} {\bibfnamefont {K.}~\bibnamefont {Yeter-Aydeniz}}, \bibinfo {author} {\bibfnamefont {R.~C.}\ \bibnamefont {Pooser}},\ and\ \bibinfo {author} {\bibfnamefont {G.}~\bibnamefont {Siopsis}},\ }\bibfield  {title} {\bibinfo {title} {Practical quantum computation of chemical and nuclear energy levels using quantum imaginary time evolution and {Lanczos} algorithms},\ }\href {https://doi.org/10.1038/s41534-020-00290-1} {\bibfield  {journal} {\bibinfo  {journal} {NPJ Quantum Inf.}\ }\textbf {\bibinfo {volume} {6}},\ \bibinfo {pages} {1} (\bibinfo {year} {2020})}\BibitemShut {NoStop}%
\bibitem [{\citenamefont {Preskill}(2018)}]{preskill_quantum_2018}%
  \BibitemOpen
  \bibfield  {author} {\bibinfo {author} {\bibfnamefont {J.}~\bibnamefont {Preskill}},\ }\bibfield  {title} {\bibinfo {title} {Quantum {Computing} in the {NISQ} era and beyond},\ }\href {https://doi.org/10.22331/q-2018-08-06-79} {\bibfield  {journal} {\bibinfo  {journal} {Quantum}\ }\textbf {\bibinfo {volume} {2}},\ \bibinfo {pages} {79} (\bibinfo {year} {2018})}\BibitemShut {NoStop}%
\bibitem [{\citenamefont {Campbell}(2021)}]{campbell_early_2021}%
  \BibitemOpen
  \bibfield  {author} {\bibinfo {author} {\bibfnamefont {E.~T.}\ \bibnamefont {Campbell}},\ }\bibfield  {title} {\bibinfo {title} {Early fault-tolerant simulations of the {Hubbard} model},\ }\href {https://doi.org/10.1088/2058-9565/ac3110} {\bibfield  {journal} {\bibinfo  {journal} {Quantum Sci. Technol.}\ }\textbf {\bibinfo {volume} {7}},\ \bibinfo {pages} {015007} (\bibinfo {year} {2022})}\BibitemShut {NoStop}%
\bibitem [{\citenamefont {Katabarwa}\ \emph {et~al.}(2024)\citenamefont {Katabarwa}, \citenamefont {Gratsea}, \citenamefont {Caesura},\ and\ \citenamefont {Johnson}}]{katabarwa_early_2024}%
  \BibitemOpen
  \bibfield  {author} {\bibinfo {author} {\bibfnamefont {A.}~\bibnamefont {Katabarwa}}, \bibinfo {author} {\bibfnamefont {K.}~\bibnamefont {Gratsea}}, \bibinfo {author} {\bibfnamefont {A.}~\bibnamefont {Caesura}},\ and\ \bibinfo {author} {\bibfnamefont {P.~D.}\ \bibnamefont {Johnson}},\ }\bibfield  {title} {\bibinfo {title} {Early {Fault}-{Tolerant} {Quantum} {Computing}},\ }\href {https://doi.org/10.1103/PRXQuantum.5.020101} {\bibfield  {journal} {\bibinfo  {journal} {PRX Quantum}\ }\textbf {\bibinfo {volume} {5}},\ \bibinfo {pages} {020101} (\bibinfo {year} {2024})}\BibitemShut {NoStop}%
\bibitem [{\citenamefont {Epperly}\ \emph {et~al.}(2022)\citenamefont {Epperly}, \citenamefont {Lin},\ and\ \citenamefont {Nakatsukasa}}]{epperly_theory_2022}%
  \BibitemOpen
  \bibfield  {author} {\bibinfo {author} {\bibfnamefont {E.~N.}\ \bibnamefont {Epperly}}, \bibinfo {author} {\bibfnamefont {L.}~\bibnamefont {Lin}},\ and\ \bibinfo {author} {\bibfnamefont {Y.}~\bibnamefont {Nakatsukasa}},\ }\bibfield  {title} {\bibinfo {title} {A theory of quantum subspace diagonalization},\ }\href {https://doi.org/10.1137/21M145954X} {\bibfield  {journal} {\bibinfo  {journal} {SIAM J. Matrix Anal. Appl.}\ }\textbf {\bibinfo {volume} {43}},\ \bibinfo {pages} {1263} (\bibinfo {year} {2022})}\BibitemShut {NoStop}%
\bibitem [{\citenamefont {Motta}\ \emph {et~al.}(2024)\citenamefont {Motta}, \citenamefont {Kirby}, \citenamefont {Liepuoniute}, \citenamefont {Sung}, \citenamefont {Cohn}, \citenamefont {Mezzacapo}, \citenamefont {Klymko}, \citenamefont {Nguyen}, \citenamefont {Yoshioka},\ and\ \citenamefont {Rice}}]{motta_subspace_2024}%
  \BibitemOpen
  \bibfield  {author} {\bibinfo {author} {\bibfnamefont {M.}~\bibnamefont {Motta}}, \bibinfo {author} {\bibfnamefont {W.}~\bibnamefont {Kirby}}, \bibinfo {author} {\bibfnamefont {I.}~\bibnamefont {Liepuoniute}}, \bibinfo {author} {\bibfnamefont {K.~J.}\ \bibnamefont {Sung}}, \bibinfo {author} {\bibfnamefont {J.}~\bibnamefont {Cohn}}, \bibinfo {author} {\bibfnamefont {A.}~\bibnamefont {Mezzacapo}}, \bibinfo {author} {\bibfnamefont {K.}~\bibnamefont {Klymko}}, \bibinfo {author} {\bibfnamefont {N.}~\bibnamefont {Nguyen}}, \bibinfo {author} {\bibfnamefont {N.}~\bibnamefont {Yoshioka}},\ and\ \bibinfo {author} {\bibfnamefont {J.~E.}\ \bibnamefont {Rice}},\ }\bibfield  {title} {\bibinfo {title} {Subspace methods for electronic structure simulations on quantum computers},\ }\href {https://doi.org/10.1088/2516-1075/ad3592} {\bibfield  {journal} {\bibinfo  {journal} {Electron. Struct.}\ }\textbf {\bibinfo {volume} {6}},\ \bibinfo {pages} {013001} (\bibinfo {year} {2024})}\BibitemShut {NoStop}%
\bibitem [{\citenamefont {Lee}\ \emph {et~al.}(2024)\citenamefont {Lee}, \citenamefont {Lee},\ and\ \citenamefont {Huh}}]{lee_sampling_2024}%
  \BibitemOpen
  \bibfield  {author} {\bibinfo {author} {\bibfnamefont {G.}~\bibnamefont {Lee}}, \bibinfo {author} {\bibfnamefont {D.}~\bibnamefont {Lee}},\ and\ \bibinfo {author} {\bibfnamefont {J.}~\bibnamefont {Huh}},\ }\bibfield  {title} {\bibinfo {title} {Sampling {Error} {Analysis} in {Quantum} {Krylov} {Subspace} {Diagonalization}},\ }\href {https://doi.org/10.22331/q-2024-09-19-1477} {\bibfield  {journal} {\bibinfo  {journal} {Quantum}\ }\textbf {\bibinfo {volume} {8}},\ \bibinfo {pages} {1477} (\bibinfo {year} {2024})}\BibitemShut {NoStop}%
\bibitem [{\citenamefont {Kirby}\ \emph {et~al.}(2023)\citenamefont {Kirby}, \citenamefont {Motta},\ and\ \citenamefont {Mezzacapo}}]{kirby_exact_2023}%
  \BibitemOpen
  \bibfield  {author} {\bibinfo {author} {\bibfnamefont {W.}~\bibnamefont {Kirby}}, \bibinfo {author} {\bibfnamefont {M.}~\bibnamefont {Motta}},\ and\ \bibinfo {author} {\bibfnamefont {A.}~\bibnamefont {Mezzacapo}},\ }\bibfield  {title} {\bibinfo {title} {Exact and efficient {Lanczos} method on a quantum computer},\ }\href {https://doi.org/10.22331/q-2023-05-23-1018} {\bibfield  {journal} {\bibinfo  {journal} {Quantum}\ }\textbf {\bibinfo {volume} {7}},\ \bibinfo {pages} {1018} (\bibinfo {year} {2023})}\BibitemShut {NoStop}%
\bibitem [{\citenamefont {Ghojogh}\ \emph {et~al.}(2023)\citenamefont {Ghojogh}, \citenamefont {Karray},\ and\ \citenamefont {Crowley}}]{ghojogh_eigenvalue_2023}%
  \BibitemOpen
  \bibfield  {author} {\bibinfo {author} {\bibfnamefont {B.}~\bibnamefont {Ghojogh}}, \bibinfo {author} {\bibfnamefont {F.}~\bibnamefont {Karray}},\ and\ \bibinfo {author} {\bibfnamefont {M.}~\bibnamefont {Crowley}},\ }\href {https://doi.org/10.48550/arXiv.1903.11240} {\bibinfo {title} {Eigenvalue and {Generalized} {Eigenvalue} {Problems}: {Tutorial}}} (\bibinfo {year} {2023}),\ \bibinfo {note} {10.48550/arXiv.1903.11240}\BibitemShut {NoStop}%
\bibitem [{\citenamefont {Parlett}(1998)}]{parlett_15_1998}%
  \BibitemOpen
  \bibfield  {author} {\bibinfo {author} {\bibfnamefont {B.~N.}\ \bibnamefont {Parlett}},\ }\bibfield  {title} {\bibinfo {title} {15. {The} {General} {Linear} {Eigenvalue} {Problem}},\ }in\ \href {https://doi.org/10.1137/1.9781611971163.ch15} {\emph {\bibinfo {booktitle} {The {Symmetric} {Eigenvalue} {Problem}}}},\ \bibinfo {series and number} {Classics in {Applied} {Mathematics}}\ (\bibinfo  {publisher} {Society for Industrial and Applied Mathematics},\ \bibinfo {year} {1998})\ pp.\ \bibinfo {pages} {339--368}\BibitemShut {NoStop}%
\bibitem [{\citenamefont {Stair}\ \emph {et~al.}(2023)\citenamefont {Stair}, \citenamefont {Cortes}, \citenamefont {Parrish}, \citenamefont {Cohn},\ and\ \citenamefont {Motta}}]{stair23_stochastic}%
  \BibitemOpen
  \bibfield  {author} {\bibinfo {author} {\bibfnamefont {N.~H.}\ \bibnamefont {Stair}}, \bibinfo {author} {\bibfnamefont {C.~L.}\ \bibnamefont {Cortes}}, \bibinfo {author} {\bibfnamefont {R.~M.}\ \bibnamefont {Parrish}}, \bibinfo {author} {\bibfnamefont {J.}~\bibnamefont {Cohn}},\ and\ \bibinfo {author} {\bibfnamefont {M.}~\bibnamefont {Motta}},\ }\bibfield  {title} {\bibinfo {title} {Stochastic quantum krylov protocol with double-factorized hamiltonians},\ }\href {https://doi.org/10.1103/PhysRevA.107.032414} {\bibfield  {journal} {\bibinfo  {journal} {Phys. Rev. A}\ }\textbf {\bibinfo {volume} {107}},\ \bibinfo {pages} {032414} (\bibinfo {year} {2023})}\BibitemShut {NoStop}%
\bibitem [{\citenamefont {Yang}\ \emph {et~al.}(2024)\citenamefont {Yang}, \citenamefont {Christianen}, \citenamefont {Bañuls}, \citenamefont {Wild},\ and\ \citenamefont {Cirac}}]{yang_phase-sensitive_2024}%
  \BibitemOpen
  \bibfield  {author} {\bibinfo {author} {\bibfnamefont {Y.}~\bibnamefont {Yang}}, \bibinfo {author} {\bibfnamefont {A.}~\bibnamefont {Christianen}}, \bibinfo {author} {\bibfnamefont {M.~C.}\ \bibnamefont {Bañuls}}, \bibinfo {author} {\bibfnamefont {D.~S.}\ \bibnamefont {Wild}},\ and\ \bibinfo {author} {\bibfnamefont {J.~I.}\ \bibnamefont {Cirac}},\ }\bibfield  {title} {\bibinfo {title} {Phase-{Sensitive} {Quantum} {Measurement} without {Controlled} {Operations}},\ }\href {https://doi.org/10.1103/PhysRevLett.132.220601} {\bibfield  {journal} {\bibinfo  {journal} {Phys. Rev. Lett.}\ }\textbf {\bibinfo {volume} {132}},\ \bibinfo {pages} {220601} (\bibinfo {year} {2024})}\BibitemShut {NoStop}%
\bibitem [{\citenamefont {Yoshioka}\ \emph {et~al.}(2025)\citenamefont {Yoshioka}, \citenamefont {Amico}, \citenamefont {Kirby}, \citenamefont {Jurcevic}, \citenamefont {Dutt}, \citenamefont {Fuller}, \citenamefont {Garion}, \citenamefont {Haas}, \citenamefont {Hamamura}, \citenamefont {Ivrii}, \citenamefont {Majumdar}, \citenamefont {Minev}, \citenamefont {Motta}, \citenamefont {Pokharel}, \citenamefont {Rivero}, \citenamefont {Sharma}, \citenamefont {Wood}, \citenamefont {Javadi-Abhari},\ and\ \citenamefont {Mezzacapo}}]{yoshioka_krylov_2025}%
  \BibitemOpen
  \bibfield  {author} {\bibinfo {author} {\bibfnamefont {N.}~\bibnamefont {Yoshioka}}, \bibinfo {author} {\bibfnamefont {M.}~\bibnamefont {Amico}}, \bibinfo {author} {\bibfnamefont {W.}~\bibnamefont {Kirby}}, \bibinfo {author} {\bibfnamefont {P.}~\bibnamefont {Jurcevic}}, \bibinfo {author} {\bibfnamefont {A.}~\bibnamefont {Dutt}}, \bibinfo {author} {\bibfnamefont {B.}~\bibnamefont {Fuller}}, \bibinfo {author} {\bibfnamefont {S.}~\bibnamefont {Garion}}, \bibinfo {author} {\bibfnamefont {H.}~\bibnamefont {Haas}}, \bibinfo {author} {\bibfnamefont {I.}~\bibnamefont {Hamamura}}, \bibinfo {author} {\bibfnamefont {A.}~\bibnamefont {Ivrii}}, \bibinfo {author} {\bibfnamefont {R.}~\bibnamefont {Majumdar}}, \bibinfo {author} {\bibfnamefont {Z.}~\bibnamefont {Minev}}, \bibinfo {author} {\bibfnamefont {M.}~\bibnamefont {Motta}}, \bibinfo {author} {\bibfnamefont {B.}~\bibnamefont {Pokharel}}, \bibinfo {author} {\bibfnamefont {P.}~\bibnamefont {Rivero}}, \bibinfo {author} {\bibfnamefont {K.}~\bibnamefont {Sharma}}, \bibinfo
  {author} {\bibfnamefont {C.~J.}\ \bibnamefont {Wood}}, \bibinfo {author} {\bibfnamefont {A.}~\bibnamefont {Javadi-Abhari}},\ and\ \bibinfo {author} {\bibfnamefont {A.}~\bibnamefont {Mezzacapo}},\ }\bibfield  {title} {\bibinfo {title} {Krylov diagonalization of large many-body {Hamiltonians} on a quantum processor},\ }\href {https://doi.org/10.1038/s41467-025-59716-z} {\bibfield  {journal} {\bibinfo  {journal} {Nat. Commun.}\ }\textbf {\bibinfo {volume} {16}},\ \bibinfo {pages} {5014} (\bibinfo {year} {2025})}\BibitemShut {NoStop}%
\bibitem [{\citenamefont {Hehre}\ \emph {et~al.}(1969)\citenamefont {Hehre}, \citenamefont {Stewart},\ and\ \citenamefont {Pople}}]{hehre_self-consistent_1969}%
  \BibitemOpen
  \bibfield  {author} {\bibinfo {author} {\bibfnamefont {W.~J.}\ \bibnamefont {Hehre}}, \bibinfo {author} {\bibfnamefont {R.~F.}\ \bibnamefont {Stewart}},\ and\ \bibinfo {author} {\bibfnamefont {J.~A.}\ \bibnamefont {Pople}},\ }\bibfield  {title} {\bibinfo {title} {Self-{Consistent} {Molecular}-{Orbital} {Methods}. {I}. {Use} of {Gaussian} {Expansions} of {Slater}-{Type} {Atomic} {Orbitals}},\ }\href {https://doi.org/10.1063/1.1672392} {\bibfield  {journal} {\bibinfo  {journal} {J. Chem. Phys.}\ }\textbf {\bibinfo {volume} {51}},\ \bibinfo {pages} {2657} (\bibinfo {year} {1969})}\BibitemShut {NoStop}%
\bibitem [{\citenamefont {Heisenberg}(1928)}]{heisenberg_zur_1928}%
  \BibitemOpen
  \bibfield  {author} {\bibinfo {author} {\bibfnamefont {W.}~\bibnamefont {Heisenberg}},\ }\bibfield  {title} {\bibinfo {title} {Zur {Theorie} des {Ferromagnetismus}},\ }\href {https://doi.org/10.1007/BF01328601} {\bibfield  {journal} {\bibinfo  {journal} {Z. Phys.}\ }\textbf {\bibinfo {volume} {49}},\ \bibinfo {pages} {619} (\bibinfo {year} {1928})}\BibitemShut {NoStop}%
\bibitem [{\citenamefont {{Zoller}}\ and\ \citenamefont {{Gardiner}}(1997)}]{1997zoller}%
  \BibitemOpen
  \bibfield  {author} {\bibinfo {author} {\bibfnamefont {P.}~\bibnamefont {{Zoller}}}\ and\ \bibinfo {author} {\bibfnamefont {C.~W.}\ \bibnamefont {{Gardiner}}},\ }\bibfield  {title} {\bibinfo {title} {{Quantum Noise in Quantum Optics: the Stochastic Schr{\"o}dinger Equation}},\ }in\ \href {https://doi.org/10.48550/arXiv.quant-ph/9702030} {\emph {\bibinfo {booktitle} {Fluctuations Quantiques/Quantum Fluctuations}}},\ \bibinfo {editor} {edited by\ \bibinfo {editor} {\bibfnamefont {S.}~\bibnamefont {{Reynaud}}}, \bibinfo {editor} {\bibfnamefont {E.}~\bibnamefont {{Giacobino}}},\ and\ \bibinfo {editor} {\bibfnamefont {J.}~\bibnamefont {{Zinn-Justin}}}}\ (\bibinfo {year} {1997})\ p.~\bibinfo {pages} {79},\ \Eprint {https://arxiv.org/abs/quant-ph/9702030} {arXiv:quant-ph/9702030 [quant-ph]} \BibitemShut {NoStop}%
\bibitem [{\citenamefont {Oliv}\ \emph {et~al.}(2022)\citenamefont {Oliv}, \citenamefont {Matic}, \citenamefont {Messerer},\ and\ \citenamefont {Lorenz}}]{Oliv2022}%
  \BibitemOpen
  \bibfield  {author} {\bibinfo {author} {\bibfnamefont {M.}~\bibnamefont {Oliv}}, \bibinfo {author} {\bibfnamefont {A.}~\bibnamefont {Matic}}, \bibinfo {author} {\bibfnamefont {T.}~\bibnamefont {Messerer}},\ and\ \bibinfo {author} {\bibfnamefont {J.~M.}\ \bibnamefont {Lorenz}},\ }\href@noop {} {\bibinfo {title} {{Evaluating the impact of noise on the performance of the Variational Quantum Eigensolver}}} (\bibinfo {year} {2022}),\ \Eprint {https://arxiv.org/abs/2209.12803} {arXiv:2209.12803 [quant-ph]} \BibitemShut {NoStop}%
\bibitem [{\citenamefont {Filip}(2024)}]{filip_fighting_2024}%
  \BibitemOpen
  \bibfield  {author} {\bibinfo {author} {\bibfnamefont {M.-A.}\ \bibnamefont {Filip}},\ }\bibfield  {title} {\bibinfo {title} {Fighting {Noise} with {Noise}: {A} {Stochastic} {Projective} {Quantum} {Eigensolver}},\ }\href {https://doi.org/10.1021/acs.jctc.4c00295} {\bibfield  {journal} {\bibinfo  {journal} {JCTC}\ }\textbf {\bibinfo {volume} {20}},\ \bibinfo {pages} {5964} (\bibinfo {year} {2024})}\BibitemShut {NoStop}%
\bibitem [{\citenamefont {Singh}\ \emph {et~al.}(2023)\citenamefont {Singh}, \citenamefont {Siwach}, \citenamefont {Singh},\ and\ \citenamefont {Arumugam}}]{singh_excitations_nodate}%
  \BibitemOpen
  \bibfield  {author} {\bibinfo {author} {\bibfnamefont {N.}~\bibnamefont {Singh}}, \bibinfo {author} {\bibfnamefont {P.}~\bibnamefont {Siwach}}, \bibinfo {author} {\bibfnamefont {A.}~\bibnamefont {Singh}},\ and\ \bibinfo {author} {\bibfnamefont {P.}~\bibnamefont {Arumugam}},\ }\bibfield  {title} {\bibinfo {title} {Excitations in {Qubit} {Space} using {Single}-{Particle} {Dipole} {Operator}},\ }\href {https://inspirehep.net/literature/2763796} {\bibfield  {journal} {\bibinfo  {journal} {DAE Symp. Nucl. Phys.}\ }\textbf {\bibinfo {volume} {67}} (\bibinfo {year} {2023})}\BibitemShut {NoStop}%
\bibitem [{\citenamefont {Bonaiti}\ \emph {et~al.}(2022)\citenamefont {Bonaiti}, \citenamefont {Bacca},\ and\ \citenamefont {Hagen}}]{bonaiti_ab_2022}%
  \BibitemOpen
  \bibfield  {author} {\bibinfo {author} {\bibfnamefont {F.}~\bibnamefont {Bonaiti}}, \bibinfo {author} {\bibfnamefont {S.}~\bibnamefont {Bacca}},\ and\ \bibinfo {author} {\bibfnamefont {G.}~\bibnamefont {Hagen}},\ }\bibfield  {title} {\bibinfo {title} {\textit{{Ab} initio} coupled-cluster calculations of ground and dipole excited states in {He} 8},\ }\href {https://doi.org/10.1103/PhysRevC.105.034313} {\bibfield  {journal} {\bibinfo  {journal} {Phys. Rev. C}\ }\textbf {\bibinfo {volume} {105}},\ \bibinfo {pages} {034313} (\bibinfo {year} {2022})}\BibitemShut {NoStop}%
\bibitem [{\citenamefont {Coester}(1958)}]{COESTER19584}%
  \BibitemOpen
  \bibfield  {author} {\bibinfo {author} {\bibfnamefont {F.}~\bibnamefont {Coester}},\ }\bibfield  {title} {\bibinfo {title} {Bound states of a many-particle system},\ }\href {https://doi.org/https://doi.org/10.1016/0029-5582(58)90280-3} {\bibfield  {journal} {\bibinfo  {journal} {Nucl. Phys.}\ }\textbf {\bibinfo {volume} {7}},\ \bibinfo {pages} {421} (\bibinfo {year} {1958})}\BibitemShut {NoStop}%
\bibitem [{\citenamefont {Bartlett}\ and\ \citenamefont {Musia\l{}}(2007)}]{CCBartlett}%
  \BibitemOpen
  \bibfield  {author} {\bibinfo {author} {\bibfnamefont {R.~J.}\ \bibnamefont {Bartlett}}\ and\ \bibinfo {author} {\bibfnamefont {M.}~\bibnamefont {Musia\l{}}},\ }\bibfield  {title} {\bibinfo {title} {Coupled-cluster theory in quantum chemistry},\ }\href {https://doi.org/10.1103/RevModPhys.79.291} {\bibfield  {journal} {\bibinfo  {journal} {Rev. Mod. Phys.}\ }\textbf {\bibinfo {volume} {79}},\ \bibinfo {pages} {291} (\bibinfo {year} {2007})}\BibitemShut {NoStop}%
\bibitem [{\citenamefont {Malz}\ \emph {et~al.}(2024)\citenamefont {Malz}, \citenamefont {Styliaris}, \citenamefont {Wei},\ and\ \citenamefont {Cirac}}]{Malz_2024_MPSprep}%
  \BibitemOpen
  \bibfield  {author} {\bibinfo {author} {\bibfnamefont {D.}~\bibnamefont {Malz}}, \bibinfo {author} {\bibfnamefont {G.}~\bibnamefont {Styliaris}}, \bibinfo {author} {\bibfnamefont {Z.-Y.}\ \bibnamefont {Wei}},\ and\ \bibinfo {author} {\bibfnamefont {J.~I.}\ \bibnamefont {Cirac}},\ }\bibfield  {title} {\bibinfo {title} {{Preparation of Matrix Product States with Log-Depth Quantum Circuits}},\ }\href {https://doi.org/10.1103/PhysRevLett.132.040404} {\bibfield  {journal} {\bibinfo  {journal} {Phys. Rev. Lett.}\ }\textbf {\bibinfo {volume} {132}},\ \bibinfo {pages} {040404} (\bibinfo {year} {2024})}\BibitemShut {NoStop}%
\bibitem [{\citenamefont {Smith}\ \emph {et~al.}(2024)\citenamefont {Smith}, \citenamefont {Khan}, \citenamefont {Clark}, \citenamefont {Girvin},\ and\ \citenamefont {Wei}}]{Smith_2024_MPSprep}%
  \BibitemOpen
  \bibfield  {author} {\bibinfo {author} {\bibfnamefont {K.~C.}\ \bibnamefont {Smith}}, \bibinfo {author} {\bibfnamefont {A.}~\bibnamefont {Khan}}, \bibinfo {author} {\bibfnamefont {B.~K.}\ \bibnamefont {Clark}}, \bibinfo {author} {\bibfnamefont {S.}~\bibnamefont {Girvin}},\ and\ \bibinfo {author} {\bibfnamefont {T.-C.}\ \bibnamefont {Wei}},\ }\bibfield  {title} {\bibinfo {title} {{Constant-Depth Preparation of Matrix Product States with Adaptive Quantum Circuits}},\ }\href {https://doi.org/10.1103/PRXQuantum.5.030344} {\bibfield  {journal} {\bibinfo  {journal} {PRX Quantum}\ }\textbf {\bibinfo {volume} {5}},\ \bibinfo {pages} {030344} (\bibinfo {year} {2024})}\BibitemShut {NoStop}%
\bibitem [{\citenamefont {Javadi-Abhari}\ \emph {et~al.}(2024)\citenamefont {Javadi-Abhari}, \citenamefont {Treinish}, \citenamefont {Krsulich}, \citenamefont {Wood}, \citenamefont {Lishman}, \citenamefont {Gacon}, \citenamefont {Martiel}, \citenamefont {Nation}, \citenamefont {Bishop}, \citenamefont {Cross}, \citenamefont {Johnson},\ and\ \citenamefont {Gambetta}}]{qiskit2024}%
  \BibitemOpen
  \bibfield  {author} {\bibinfo {author} {\bibfnamefont {A.}~\bibnamefont {Javadi-Abhari}}, \bibinfo {author} {\bibfnamefont {M.}~\bibnamefont {Treinish}}, \bibinfo {author} {\bibfnamefont {K.}~\bibnamefont {Krsulich}}, \bibinfo {author} {\bibfnamefont {C.~J.}\ \bibnamefont {Wood}}, \bibinfo {author} {\bibfnamefont {J.}~\bibnamefont {Lishman}}, \bibinfo {author} {\bibfnamefont {J.}~\bibnamefont {Gacon}}, \bibinfo {author} {\bibfnamefont {S.}~\bibnamefont {Martiel}}, \bibinfo {author} {\bibfnamefont {P.~D.}\ \bibnamefont {Nation}}, \bibinfo {author} {\bibfnamefont {L.~S.}\ \bibnamefont {Bishop}}, \bibinfo {author} {\bibfnamefont {A.~W.}\ \bibnamefont {Cross}}, \bibinfo {author} {\bibfnamefont {B.~R.}\ \bibnamefont {Johnson}},\ and\ \bibinfo {author} {\bibfnamefont {J.~M.}\ \bibnamefont {Gambetta}},\ }\href {https://doi.org/10.48550/arXiv.2405.08810} {\bibinfo {title} {Quantum computing with {Q}iskit}} (\bibinfo {year} {2024}),\ \Eprint {https://arxiv.org/abs/2405.08810} {arXiv:2405.08810} \BibitemShut
  {NoStop}%
\bibitem [{\citenamefont {Hatano}\ and\ \citenamefont {Suzuki}(2005)}]{hatano_2005}%
  \BibitemOpen
  \bibfield  {author} {\bibinfo {author} {\bibfnamefont {N.}~\bibnamefont {Hatano}}\ and\ \bibinfo {author} {\bibfnamefont {M.}~\bibnamefont {Suzuki}},\ }\bibinfo {title} {Finding exponential product formulas of higher orders},\ in\ \href {https://doi.org/10.1007/11526216_2} {\emph {\bibinfo {booktitle} {Quantum Annealing and Other Optimization Methods}}}\ (\bibinfo  {publisher} {Springer Berlin Heidelberg},\ \bibinfo {year} {2005})\ p.\ \bibinfo {pages} {37–68}\BibitemShut {NoStop}%
\bibitem [{\citenamefont {Berry}\ \emph {et~al.}(2007)\citenamefont {Berry}, \citenamefont {Ahokas}, \citenamefont {Cleve},\ and\ \citenamefont {Sanders}}]{berry_efficient_2007}%
  \BibitemOpen
  \bibfield  {author} {\bibinfo {author} {\bibfnamefont {D.~W.}\ \bibnamefont {Berry}}, \bibinfo {author} {\bibfnamefont {G.}~\bibnamefont {Ahokas}}, \bibinfo {author} {\bibfnamefont {R.}~\bibnamefont {Cleve}},\ and\ \bibinfo {author} {\bibfnamefont {B.~C.}\ \bibnamefont {Sanders}},\ }\bibfield  {title} {\bibinfo {title} {Efficient quantum algorithms for simulating sparse {Hamiltonians}},\ }\href {https://doi.org/10.1007/s00220-006-0150-x} {\bibfield  {journal} {\bibinfo  {journal} {Commun. Math. Phys.}\ }\textbf {\bibinfo {volume} {270}},\ \bibinfo {pages} {359} (\bibinfo {year} {2007})}\BibitemShut {NoStop}%
\bibitem [{\citenamefont {Lee}\ \emph {et~al.}(2025)\citenamefont {Lee}, \citenamefont {Choi}, \citenamefont {Huh},\ and\ \citenamefont {Izmaylov}}]{lee_efficient_2025}%
  \BibitemOpen
  \bibfield  {author} {\bibinfo {author} {\bibfnamefont {G.}~\bibnamefont {Lee}}, \bibinfo {author} {\bibfnamefont {S.}~\bibnamefont {Choi}}, \bibinfo {author} {\bibfnamefont {J.}~\bibnamefont {Huh}},\ and\ \bibinfo {author} {\bibfnamefont {A.~F.}\ \bibnamefont {Izmaylov}},\ }\bibfield  {title} {\bibinfo {title} {Efficient {Strategies} for {Reducing} {Sampling} {Error} in {Quantum} {Krylov} {Subspace} {Diagonalization}},\ }\href {https://doi.org/10.1039/D4DD00321G} {\bibfield  {journal} {\bibinfo  {journal} {Digit. Discov.}\ ,\ \bibinfo {pages} {954}} (\bibinfo {year} {2025})}\BibitemShut {NoStop}%
\bibitem [{\citenamefont {Kirby}(2024)}]{Kirby_2024}%
  \BibitemOpen
  \bibfield  {author} {\bibinfo {author} {\bibfnamefont {W.}~\bibnamefont {Kirby}},\ }\bibfield  {title} {\bibinfo {title} {Analysis of quantum krylov algorithms with errors},\ }\href {https://doi.org/10.22331/q-2024-08-29-1457} {\bibfield  {journal} {\bibinfo  {journal} {Quantum}\ }\textbf {\bibinfo {volume} {8}},\ \bibinfo {pages} {1457} (\bibinfo {year} {2024})}\BibitemShut {NoStop}%
\bibitem{OliveiraGlaser2025}
Maria Gabriela Oliveira and Nina Glaser,
\textit{NQCP/NQCP-AA-pub-QBKSP\_eigensolver: Release 1.0.0 (v1.0.0)},
Zenodo, 2025.
\url{https://doi.org/10.5281/zenodo.17552224}\BibitemShut {NoStop}%
\end{thebibliography}

%

\appendix

\section{Quantum circuits proofs}
\label{app:quantumc}
This appendix shows the equivalence of the three quantum circuits in \autoref{H circuit} to evaluate the desired expectation values. 
For all three cases below note that we define $\ket{\psi^\alpha}=W_\alpha \ket{\mathbf{0}}$ and $\ket{\psi^\beta}=W_\beta \ket{\mathbf{0}}$, where $\ket{\mathbf{0}}$ corresponds to all qubits in state $\ket{0}$, and $\ket{\psi^\beta}=V_{\beta\alpha} \ket{\psi^\alpha}$.

\subsection{Quantum circuit (a)}

\begin{figure}[H]
 \centering
\includegraphics*{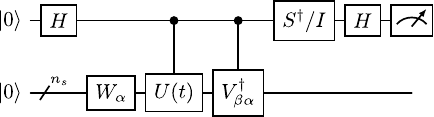}
\end{figure}

Applying the quantum gates in the circuit above for the real part, step-by-step, leads to 

\begin{align}
    \ket{0}\ket{\mathbf{0}} &\xrightarrow{H\otimes I} \frac{1}{\sqrt{2}}\left(\ket{0}+\ket{1}\right)\ket{\mathbf{0}} \nonumber \\
    &\xrightarrow{I\otimes W_\alpha} \frac{1}{\sqrt{2}}\left(\ket{0}+\ket{1}\right)\ket{\psi^\alpha} \nonumber \\
    &\xrightarrow{C_{U(t)}}\frac{1}{\sqrt{2}} \left(\ket{0}\ket{\psi^\alpha} + \ket{1}U(t)\ket{\psi^\alpha}\right)\nonumber \\
    &\xrightarrow{C_{V^\dagger_{\beta\alpha}}}\frac{1}{\sqrt{2}} \left(\ket{0}\ket{\psi^\alpha} + \ket{1}V^\dagger_{\beta\alpha} U(t)\ket{\psi^\alpha}\right)\nonumber \\
    &\xrightarrow{H\otimes I} \frac{1}{2}\left(\ket{0}\ket{\psi^\alpha} +\ket{1}\ket{\psi^\alpha} +\ket{0}V^\dagger_{\beta\alpha} U(t)\ket{\psi^\alpha} \right. \nonumber \\
    &\quad \quad \quad \quad \left. - \ket{1}V^\dagger_{\beta\alpha} U(t)\ket{\psi^\alpha}\right) \nonumber \\
    &= \frac{1}{2}\left(\ket{0} \left(\ket{\psi^\alpha} + V^\dagger_{\beta\alpha} U(t)\ket{\psi^\alpha}\right) \right. \nonumber \\
    &\quad \left. +\ket{1}\left(\ket{\psi^\alpha}  - V^\dagger_{\beta\alpha} U(t)\ket{\psi^\alpha}\right)\right).
\end{align}

The probability of measuring the ancilla qubit in state $\ket{0}$ in the above final state is 
\begin{widetext}
\begin{align}
     P(\ket{0})&= \frac{1}{4} \left(\bra{\psi^\alpha} + \bra{\psi^\alpha}\left(V^\dagger_{\beta\alpha} U(t)\right)^\dagger\right)\left(\ket{\psi^\alpha} + V^\dagger_{\beta\alpha} U(t)\ket{\psi^\alpha}\right) \nonumber \\
    &=\frac{1}{4} \left(\bra{\psi^\alpha}\psi^\alpha \rangle + \bra{\psi^\alpha}\left(V^\dagger_{\beta\alpha} U(t)\right)^\dagger\ket{\psi^\alpha} + \bra{\psi^\alpha}V^\dagger_{\beta\alpha} U(t)\ket{\psi^\alpha} + \bra{\psi^\alpha}\left(V^\dagger_{\beta\alpha} U(t)\right)^\dagger V^\dagger_{\beta\alpha} U(t)\ket{\psi^\alpha}\right) \nonumber \\
    &=\frac{1}{4} \left(1 + \left(\bra{\psi^\beta}U(t)\ket{\psi^\alpha}\right)^* + \bra{\psi^\beta} U(t)\ket{\psi^\alpha} + 1\right) = \frac{1}{2} \left(1 + \Re\left(\bra{\psi^\beta} U(t)\ket{\psi^\alpha}\right)\right).
\end{align}
\end{widetext}
Similarly, the probability of measuring the ancilla qubit in state $\ket{1}$ is 
\begin{widetext}

\begin{align}
    P(\ket{1})&= \frac{1}{4} \left(\bra{\psi^\alpha} - \bra{\psi^\alpha}\left(V^\dagger_{\beta\alpha} U(t)\right)^\dagger\right)\left(\ket{\psi^\alpha} - V^\dagger_{\beta\alpha} U(t)\ket{\psi^\alpha}\right) \nonumber \\
    &=\frac{1}{4} \left(\bra{\psi^\alpha}\psi^\alpha \rangle - \bra{\psi^\alpha}\left(V^\dagger_{\beta\alpha} U(t)\right)^\dagger\ket{\psi^\alpha} - \bra{\psi^\alpha}V^\dagger_{\beta\alpha} U(t)\ket{\psi^\alpha} + \bra{\psi^\alpha}\left(V^\dagger_{\beta\alpha} U(t)\right)^\dagger V^\dagger_{\beta\alpha} U(t)\ket{\psi^\alpha}\right) \nonumber \\
    &=\frac{1}{4} \left(1 - \left(\bra{\psi^\beta}U(t)\ket{\psi^\alpha}\right)^* - \bra{\psi^\beta} U(t)\ket{\psi^\alpha} + 1\right) = \frac{1}{2} \left(1 - \Re\left(\bra{\psi^\beta} U(t)\ket{\psi^\alpha}\right)\right).
\end{align}
\end{widetext}
Thus, $P(\ket{0})- P(\ket{1})= \Re\left(\bra{\psi^\beta} U(t)\ket{\psi^\alpha}\right)$, i.e., the real part of the desired expectation value. Doing the analogous calculation for quantum circuit concerning the imaginary part leads to 
\allowdisplaybreaks

\begin{align}
    \ket{0}\ket{\mathbf{0}} &\xrightarrow{H\otimes I} \frac{1}{\sqrt{2}}\left(\ket{0}+\ket{1}\right)\ket{\mathbf{0}} \nonumber \\
    &\xrightarrow{I\otimes W_\alpha} \frac{1}{\sqrt{2}}\left(\ket{0}+\ket{1}\right)\ket{\psi^\alpha} \nonumber \\
    &\xrightarrow{C_{U(t)}}\frac{1}{\sqrt{2}} \left(\ket{0}\ket{\psi^\alpha} + \ket{1}U(t)\ket{\psi^\alpha}\right)\nonumber \\
    &\xrightarrow{C_{V^\dagger_{\beta\alpha}}}\frac{1}{\sqrt{2}} \left(\ket{0}\ket{\psi^\alpha} + \ket{1}V^\dagger_{\beta\alpha} U(t)\ket{\psi^\alpha}\right)\nonumber \\
    &\xrightarrow{S^\dagger}\frac{1}{\sqrt{2}} \left(\ket{0}\ket{\psi^\alpha} -i \ket{1}V^\dagger_{\beta\alpha} U(t)\ket{\psi^\alpha}\right)\nonumber \\
    &\xrightarrow{H\otimes I} \frac{1}{2}\left(\ket{0}\ket{\psi^\alpha} +\ket{1}\ket{\psi^\alpha} -i\ket{0}V^\dagger_{\beta\alpha} U(t)\ket{\psi^\alpha} \right. \nonumber \\
    &\left.+i\ket{1}V^\dagger_{\beta\alpha} U(t)\ket{\psi^\alpha}\right) \nonumber \\
    &= \frac{1}{2}\left(\ket{0} \left(\ket{\psi^\alpha} -i V^\dagger_{\beta\alpha} U(t)\ket{\psi^\alpha}\right)\right. \nonumber \\
    &\left. +\ket{1}\left(\ket{\psi^\alpha}  +i V^\dagger_{\beta\alpha} U(t)\ket{\psi^\alpha}\right)\right).
\end{align}

Analogous to the real part case, the probability of measuring the ancilla qubit in state $\ket{0}$ evaluates to 
\begin{widetext}
\begin{align}
     P(\ket{0})&= \frac{1}{4} \left(\bra{\psi^\alpha} + i\bra{\psi^\alpha}\left(V^\dagger_{\beta\alpha} U(t)\right)^\dagger\right)\left(\ket{\psi^\alpha} - i V^\dagger_{\beta\alpha} U(t)\ket{\psi^\alpha}\right) \nonumber \\
    &=\frac{1}{4} \left(\bra{\psi^\alpha}\psi^\alpha \rangle + i\bra{\psi^\alpha}\left(V^\dagger_{\beta\alpha} U(t)\right)^\dagger\ket{\psi^\alpha} -i\bra{\psi^\alpha}V^\dagger_{\beta\alpha} U(t)\ket{\psi^\alpha} + \bra{\psi^\alpha}\left(V^\dagger_{\beta\alpha} U(t)\right)^\dagger V^\dagger_{\beta\alpha} U(t)\ket{\psi^\alpha}\right) \nonumber \\
    &=\frac{1}{4} \left(1 + i\left(\bra{\psi^\beta}U(t)\ket{\psi^\alpha}\right)^* -i \bra{\psi^\beta} U(t)\ket{\psi^\alpha} + 1\right) = \frac{1}{2} \left(1 + \Im\left(\bra{\psi^\beta} U(t)\ket{\psi^\alpha}\right)\right),
\end{align}
\end{widetext}
and the probability of measuring the ancilla qubit in state $\ket{1}$ is 
\begin{widetext}
\begin{align}
    P(\ket{1})&= \frac{1}{4} \left(\bra{\psi^\alpha} - i\bra{\psi^\alpha}\left(V^\dagger_{\beta\alpha} U(t)\right)^\dagger\right)\left(\ket{\psi^\alpha} + i V^\dagger_{\beta\alpha} U(t)\ket{\psi^\alpha}\right) \nonumber \\
    &=\frac{1}{4} \left(\bra{\psi^\alpha}\psi^\alpha \rangle - i\bra{\psi^\alpha}\left(V^\dagger_{\beta\alpha} U(t)\right)^\dagger\ket{\psi^\alpha} +i\bra{\psi^\alpha}V^\dagger_{\beta\alpha} U(t)\ket{\psi^\alpha} + \bra{\psi^\alpha}\left(V^\dagger_{\beta\alpha} U(t)\right)^\dagger V^\dagger_{\beta\alpha} U(t)\ket{\psi^\alpha}\right) \nonumber \\
    &=\frac{1}{4} \left(1 - i\left(\bra{\psi^\beta}U(t)\ket{\psi^\alpha}\right)^* +i \bra{\psi^\beta} U(t)\ket{\psi^\alpha} + 1\right) = \frac{1}{2} \left(1 -\Im\left(\bra{\psi^\beta} U(t)\ket{\psi^\alpha}\right)\right).
\end{align}
\end{widetext}
So, $P(\ket{0})- P(\ket{1})= \Im\left(\bra{\psi^\beta} U(t)\ket{\psi^\alpha}\right)$.

\subsection{Quantum circuit (b)}

\begin{figure}[H]
 \centering
\includegraphics*{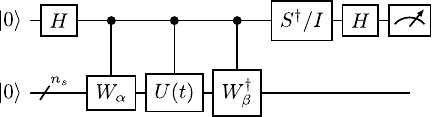}
\end{figure}

Following the same procedure as before for circuit b with $S^\dagger/I$ set to $I$, leads to the state
\begin{align}
    \ket{0}\ket{\mathbf{0}} &\xrightarrow{H\otimes I} \frac{1}{\sqrt{2}}\left(\ket{0}+\ket{1}\right)\ket{\mathbf{0}} \nonumber \\
    &\xrightarrow{C_{W_\alpha}} \frac{1}{\sqrt{2}}\left(\ket{0}\ket{\mathbf{0}}+\ket{1}\ket{\psi^\alpha}\right) \nonumber \\
    &\xrightarrow{C_{U(t)}}\frac{1}{\sqrt{2}} \left(\ket{0}\ket{\mathbf{0}} + \ket{1}U(t)\ket{\psi^\alpha}\right)\nonumber \\
    &\xrightarrow{C_{W^\dagger_\beta}}\frac{1}{\sqrt{2}} \left(\ket{0}\ket{\mathbf{0}} + \ket{1}W^\dagger_\beta U(t)\ket{\psi^\alpha}\right)\nonumber \\
    &\xrightarrow{H\otimes I} \frac{1}{2}\left(\ket{0}\ket{\mathbf{0}}+\ket{1}\ket{\mathbf{0}} +\ket{0}W^\dagger_\beta U(t)\ket{\psi^\alpha} \right. \nonumber \\
    &\left.\quad\quad\quad- \ket{1}W^\dagger_\beta U(t)\ket{\psi^\alpha}\right) \nonumber \\
    &= \frac{1}{2}\left(\ket{0} \left(\ket{\mathbf{0}} + W^\dagger_\beta U(t)\ket{\psi^\alpha}\right)\right. \nonumber \\
    &\left.+\ket{1}\left(\ket{\mathbf{0}}  - W^\dagger_\beta U(t)\ket{\psi^\alpha}\right)\right).
\end{align}

Here, the probability of measuring the ancilla qubit in state $\ket{0}$ is 
\begin{widetext}
\begin{align}
     P(\ket{0})&= \frac{1}{4} \left(\bra{\mathbf{0}} + \bra{\psi^\alpha}\left(W^\dagger_\beta U(t)\right)^\dagger\right)\left(\ket{\mathbf{0}} + W^\dagger_\beta U(t)\ket{\psi^\alpha}\right) \nonumber \\
    &=\frac{1}{4} \left(\bra{\mathbf{0}}\mathbf{0} \rangle + \bra{\psi^\alpha}\left(W^\dagger_\beta U(t)\right)^\dagger\ket{\mathbf{0}} + \bra{\mathbf{0}}W^\dagger_\beta U(t)\ket{\psi^\alpha} + \bra{\psi^\alpha}\left(W^\dagger_\beta U(t)\right)^\dagger W^\dagger_\beta U(t)\ket{\psi^\alpha}\right) \nonumber \\
    &=\frac{1}{4} \left(1 + \left(\bra{\psi^\beta}U(t)\ket{\psi^\alpha}\right)^* + \bra{\psi^\beta} U(t)\ket{\psi^\alpha} + 1\right) = \frac{1}{2} \left(1 + \Re\left(\bra{\psi^\beta} U(t)\ket{\psi^\alpha}\right)\right),
\end{align}

and the probability of measuring the ancilla qubit in state $\ket{1}$ is

\begin{align}
     P(\ket{1})&= \frac{1}{4} \left(\bra{\mathbf{0}} - \bra{\psi^\alpha}\left(W^\dagger_\beta U(t)\right)^\dagger\right)\left(\ket{\mathbf{0}} - W^\dagger_\beta U(t)\ket{\psi^\alpha}\right) \nonumber \\
    &=\frac{1}{4} \left(\bra{\mathbf{0}}\mathbf{0} \rangle - \bra{\psi^\alpha}\left(W^\dagger_\beta U(t)\right)^\dagger\ket{\mathbf{0}} - \bra{\mathbf{0}}W^\dagger_\beta U(t)\ket{\psi^\alpha} + \bra{\psi^\alpha}\left(W^\dagger_\beta U(t)\right)^\dagger W^\dagger_\beta U(t)\ket{\psi^\alpha}\right) \nonumber \\
    &=\frac{1}{4} \left(1 - \left(\bra{\psi^\beta}U(t)\ket{\psi^\alpha}\right)^* - \bra{\psi^\beta} U(t)\ket{\psi^\alpha} + 1\right) = \frac{1}{2} \left(1 - \Re\left(\bra{\psi^\beta} U(t)\ket{\psi^\alpha}\right)\right).
\end{align}
\end{widetext}
Consequently, $P(\ket{0})- P(\ket{1})= \Re\left(\bra{\psi^\beta} U(t)\ket{\psi^\alpha}\right)$. Similarly, the final state of the quantum circuit for the imaginary part case is 

\begin{align}
    \ket{0}\ket{\mathbf{0}} &\xrightarrow{H\otimes I} \frac{1}{\sqrt{2}}\left(\ket{0}+\ket{1}\right)\ket{\mathbf{0}} \nonumber \\
    &\xrightarrow{C_{W_\alpha}} \frac{1}{\sqrt{2}}\left(\ket{0}\ket{\mathbf{0}}+\ket{1}\ket{\psi^\alpha}\right) \nonumber \\
    &\xrightarrow{C_{U(t)}}\frac{1}{\sqrt{2}} \left(\ket{0}\ket{\mathbf{0}} + \ket{1}U(t)\ket{\psi^\alpha}\right)\nonumber \\
    &\xrightarrow{C_{W^\dagger_\beta}}\frac{1}{\sqrt{2}} \left(\ket{0}\ket{\mathbf{0}} + \ket{1}W^\dagger_\beta U(t)\ket{\psi^\alpha}\right)\nonumber \\
    &\xrightarrow{S^\dagger}\frac{1}{\sqrt{2}} \left(\ket{0}\ket{\mathbf{0}} -i \ket{1}W^\dagger_\beta U(t)\ket{\psi^\alpha}\right)\nonumber \\
    &\xrightarrow{H\otimes I} \frac{1}{2}\left(\ket{0}\ket{\mathbf{0}} +\ket{1}\ket{\mathbf{0}} -i\ket{0}W^\dagger_\beta U(t)\ket{\psi^\alpha} \right. \nonumber \\
    &\quad\quad\quad\left.+i\ket{1}W^\dagger_\beta U(t)\ket{\psi^\alpha}\right) \nonumber \\
    &= \frac{1}{2}\left(\ket{0} \left(\ket{\mathbf{0}} -i W^\dagger_\beta U(t)\ket{\psi^\alpha}\right)\right. \nonumber \\
    &\quad \left. +\ket{1}\left(\ket{\mathbf{0}}  +i W^\dagger_\beta U(t)\ket{\psi^\alpha}\right)\right),
\end{align}

with the probability of measuring the ancilla qubit in state $\ket{0}$ being
\begin{widetext}
\begin{align}
     P(\ket{0})&= \frac{1}{4} \left(\bra{\mathbf{0}} + i\bra{\psi^\alpha}\left(W^\dagger_\beta U(t)\right)^\dagger\right)\left(\ket{\mathbf{0}} - i W^\dagger_\beta U(t)\ket{\psi^\alpha}\right) \nonumber \\
    &=\frac{1}{4} \left(\bra{\mathbf{0}}\mathbf{0} \rangle + i\bra{\psi^\alpha}\left(W^\dagger_\beta U(t)\right)^\dagger\ket{\mathbf{0}} -i\bra{\mathbf{0}}W^\dagger_\beta U(t)\ket{\psi^\alpha} + \bra{\psi^\alpha}\left(W^\dagger_\beta U(t)\right)^\dagger W^\dagger_\beta U(t)\ket{\psi^\alpha}\right) \nonumber \\
    &=\frac{1}{4} \left(1 + i\left(\bra{\psi^\beta}U(t)\ket{\psi^\alpha}\right)^* -i \bra{\psi^\beta} U(t)\ket{\psi^\alpha} + 1\right) = \frac{1}{2} \left(1 +\Im\left(\bra{\psi^\beta} U(t)\ket{\psi^\alpha}\right)\right),
\end{align}
\end{widetext}
and in state $\ket{1}$ being
\begin{widetext}
\begin{align}
    P(\ket{1})&= \frac{1}{4} \left(\bra{\mathbf{0}} - i\bra{\psi^\alpha}\left(W^\dagger_\beta U(t)\right)^\dagger\right)\left(\ket{\mathbf{0}}+- i W^\dagger_\beta U(t)\ket{\psi^\alpha}\right) \nonumber \\
    &=\frac{1}{4} \left(\bra{\mathbf{0}}\mathbf{0} \rangle - i\bra{\psi^\alpha}\left(W^\dagger_\beta U(t)\right)^\dagger\ket{\mathbf{0}} +i\bra{\mathbf{0}}W^\dagger_\beta U(t)\ket{\psi^\alpha} + \bra{\psi^\alpha}\left(W^\dagger_\beta U(t)\right)^\dagger W^\dagger_\beta U(t)\ket{\psi^\alpha}\right) \nonumber \\
    &=\frac{1}{4} \left(1 - i\left(\bra{\psi^\beta}U(t)\ket{\psi^\alpha}\right)^* +i \bra{\psi^\beta} U(t)\ket{\psi^\alpha} + 1\right) = \frac{1}{2} \left(1 -\Im\left(\bra{\psi^\beta} U(t)\ket{\psi^\alpha}\right)\right).
\end{align}

Thus, $P(\ket{0})- P(\ket{1})= \Im\left(\bra{\psi^\beta} U(t)\ket{\psi^\alpha}\right)$.
\end{widetext}

\clearpage

\subsection{Quantum circuit (c)}

\begin{figure}[H]
 \centering
\includegraphics*{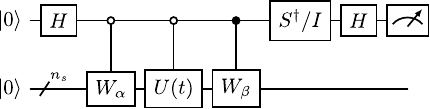}
\end{figure}

Lastly, doing the same calculation for the third circuit with $I$ leads to the state
\begin{align}
    \ket{0}\ket{\mathbf{0}} &\xrightarrow{H\otimes I} \frac{1}{\sqrt{2}}\left(\ket{0}+\ket{1}\right)\ket{\mathbf{0}} \nonumber \\
    &\xrightarrow{C^0_{W_\alpha}} \frac{1}{\sqrt{2}}\left(\ket{0}\ket{\psi^\alpha}+\ket{1}\ket{\mathbf{0}}\right) \nonumber \\
    &\xrightarrow{C^0_{U(t)}}\frac{1}{\sqrt{2}} \left(\ket{0}U(t)\ket{\psi^\alpha} + \ket{1}\ket{\mathbf{0}}\right)\nonumber \\
    &\xrightarrow{C_{W_\beta}}\frac{1}{\sqrt{2}} \left(\ket{0}U(t)\ket{\psi^\alpha} + \ket{1}\ket{\psi^\beta}\right)\nonumber \\
    &\xrightarrow{H\otimes I} \frac{1}{2}\left(\ket{0}U(t)\ket{\psi^\alpha}+\ket{1}U(t)\ket{\psi^\alpha}\right. \nonumber \\
    &\quad \quad \quad \left.+\ket{0}\ket{\psi^\beta} - \ket{1}\ket{\psi^\beta}\right) \nonumber \\
    &= \frac{1}{2}\left(\ket{0} \left(U(t)\ket{\psi^\alpha} + \ket{\psi^\beta}\right)\right. \nonumber \\
    &\left.+\ket{1}\left(U(t)\ket{\psi^\alpha}  - \ket{\psi^\beta}\right)\right).
\end{align}

Here, the probability of measuring the ancilla qubit in state $\ket{0}$ is 
\begin{widetext}
\begin{align}
     P(\ket{0})&= \frac{1}{4} \left(\bra{\psi^\alpha}U^\dagger(t) + \bra{\psi^\beta}\right)\left((U(t)\ket{\psi^\alpha}  + \ket{\psi^\beta}\right) \nonumber \\
    &=\frac{1}{4} \left(\bra{\psi^\alpha}U^\dagger(t)U(t)\ket{\psi^\alpha} + \bra{\psi^\alpha}U^\dagger(t)\ket{\psi^\beta} + \bra{\psi^\beta} U(t)\ket{\psi^\alpha} + \bra{\psi^\beta}\psi^\beta \rangle\right) \nonumber \\
    &=\frac{1}{4} \left(1 + \left(\bra{\psi^\beta}U(t)\ket{\psi^\alpha}\right)^* + \bra{\psi^\beta} U(t)\ket{\psi^\alpha} + 1\right) = \frac{1}{2} \left(1 + \Re\left(\bra{\psi^\beta} U(t)\ket{\psi^\alpha}\right)\right).
\end{align}
\end{widetext}
and the one of measuring the ancilla qubit in state $\ket{1}$ is 
\begin{align}
     P(\ket{1})&=  \frac{1}{2} \left(1 - \Re\left(\bra{\psi^\beta} U(t)\ket{\psi^\alpha}\right)\right).
\end{align}

Thus, $P(\ket{0})- P(\ket{1})= \Re\left(\bra{\psi^\beta} U(t)\ket{\psi^\alpha}\right)$.
The imaginary part of this circuit corresponds to the final state
\begin{align}
    \ket{0}\ket{\mathbf{0}} &\xrightarrow{H\otimes I} \frac{1}{\sqrt{2}}\left(\ket{0}+\ket{1}\right)\ket{\mathbf{0}} \nonumber \\
    &\xrightarrow{C^0_{W_\alpha}} \frac{1}{\sqrt{2}}\left(\ket{0}\ket{\psi^\alpha}+\ket{1}\ket{\mathbf{0}}\right) \nonumber \\
    &\xrightarrow{C^0_{U(t)}}\frac{1}{\sqrt{2}} \left(\ket{0}U(t)\ket{\psi^\alpha} + \ket{1}\ket{\mathbf{0}}\right)\nonumber \\
    &\xrightarrow{C_{W_\beta}}\frac{1}{\sqrt{2}} \left(\ket{0}U(t)\ket{\psi^\alpha} + \ket{1}\ket{\psi^\beta}\right)\nonumber \\
    &\xrightarrow{S^\dagger}\frac{1}{\sqrt{2}} \left(\ket{0}U(t)\ket{\psi^\alpha} -i \ket{1}\ket{\psi^\beta}\right)\nonumber \\
    &\xrightarrow{H\otimes I} \frac{1}{2}\left(\ket{0}U(t)\ket{\psi^\alpha}+\ket{1}U(t)\ket{\psi^\alpha}\right. \nonumber \\
    &\left.-i\ket{0}\ket{\psi^\beta} +i \ket{1}\ket{\psi^\beta}\right) \nonumber \\
    &= \frac{1}{2}\left(\ket{0} \left(U(t)\ket{\psi^\alpha} -i \ket{\psi^\beta}\right)\right. \nonumber \\
    &\left.+\ket{1}\left(U(t)\ket{\psi^\alpha}  +i \ket{\psi^\beta}\right)\right),
\end{align}

with the probability of measuring the ancilla qubit in state $\ket{0}$ being 
\begin{widetext}
\begin{align}
     P(\ket{0})&= \frac{1}{4} \left(\bra{\psi^\alpha}U^\dagger(t) + i\bra{\psi^\beta}\right)\left((U(t)\ket{\psi^\alpha}  -i \ket{\psi^\beta}\right) \nonumber \\
    &=\frac{1}{4} \left(\bra{\psi^\alpha}U^\dagger(t)U(t)\ket{\psi^\alpha} -i \bra{\psi^\alpha}U^\dagger(t)\ket{\psi^\beta} + i\bra{\psi^\beta} U(t)\ket{\psi^\alpha} + \bra{\psi^\beta}\psi^\beta \rangle\right) \nonumber \\
    &=\frac{1}{4} \left(1 -i \left(\bra{\psi^\beta}U(t)\ket{\psi^\alpha}\right)^* + i\bra{\psi^\beta} U(t)\ket{\psi^\alpha} + 1\right) = \frac{1}{2} \left(1 - \Im\left(\bra{\psi^\beta} U(t)\ket{\psi^\alpha}\right)\right),
\end{align}
\end{widetext}
and the one of measuring the ancilla qubit in state $\ket{1}$
\begin{align}
    P(\ket{1})& = \frac{1}{2} \left(1 + \Im\left(\bra{\psi^\beta} U(t)\ket{\psi^\alpha}\right)\right)
\end{align}

Consequently, $P(\ket{1})- P(\ket{0})= \Im\left(\bra{\psi^\beta} U(t)\ket{\psi^\alpha}\right)$.

\section{Overlap matrix regularisation threshold}
\label{app:Sthre}

\begin{figure*}
    \centering
    \includegraphics[width=1\linewidth]{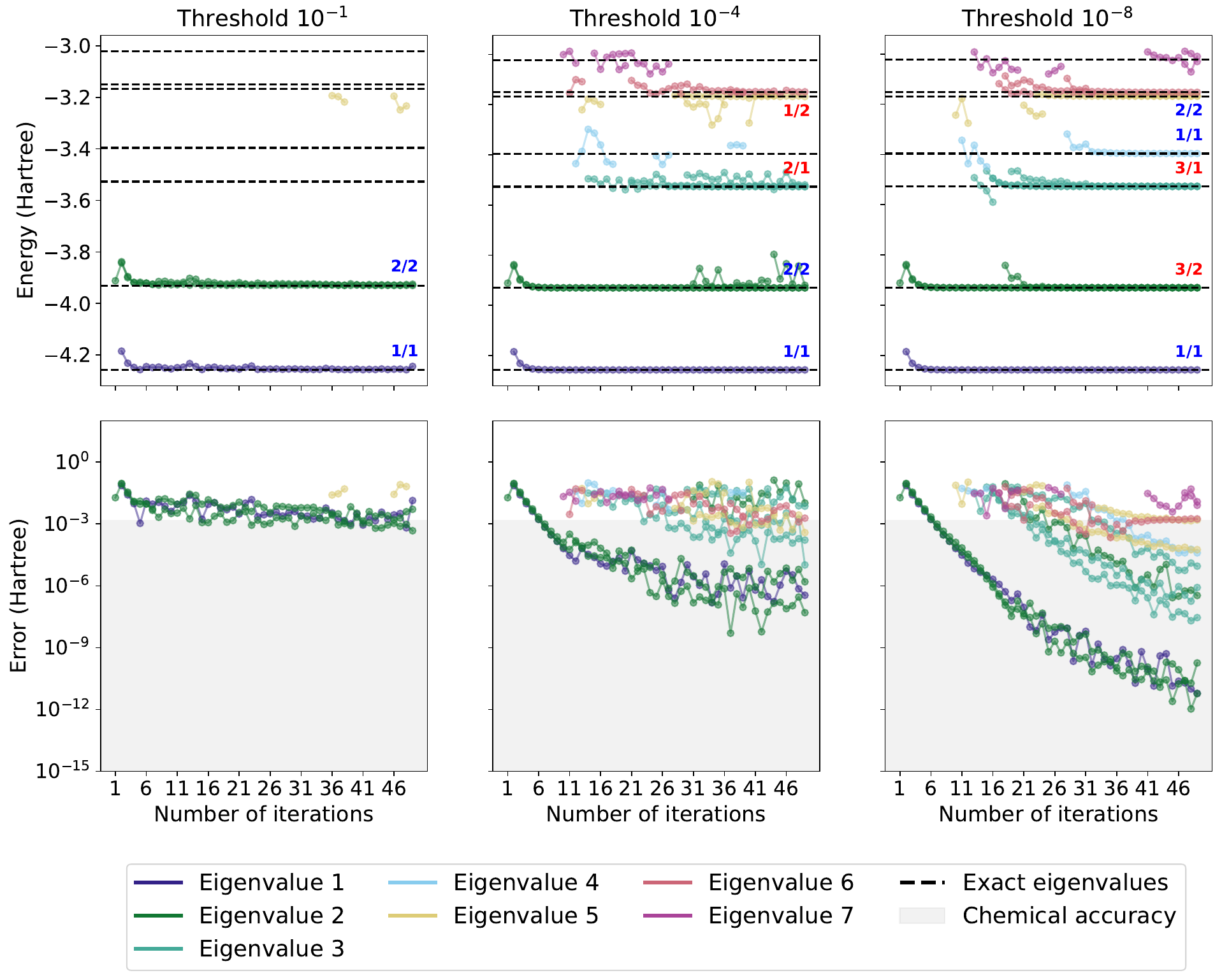}
    \caption{QBKSP convergence for the 10-site Heisenberg model, evaluated for different regularization threshold values for three initial states with $0.5$ overlap, and a fixed evolution time of $\tau=3$ atu. 
    The top panel shows the convergence of the seven lowest eigenvalues as a function of the number of iterations, whereas the bottom panel shows the absolute error of the calculated eigenvalues.
    The multiplicity of degenerate energies found for each eigenvalue after 50 iterations and the correct multiplicity of that eigenvalue are denoted next to the energies.
    The blue colour indicates that the right multiplicity was found, whereas the red one indicates that the retrieved degeneracies do not correspond to the actual multiplicity of the state. It should be noted that in these calculations, no convergence criterion was employed, such that spurious states caused by emerging linear dependencies can arise over time.
   }
    \label{fig:10Heisenberginitial_s4}
\end{figure*}

In \autoref{fig:10Heisenberginitial_s4}, we study the impact of the regularization threshold parameter on the QBKSP convergence for the 10-site Heisenberg model. 
In ideal numerical simulations, smaller threshold values yield higher accuracy and enable the retrieval of more eigenvalues. 
However, when the calculation is converged for several iterations, the observed degeneracy multiplicities can exceed the expected values. In contrast, larger threshold values limit the achievable accuracy but correctly reproduce the expected number of degenerate states.

\section{LiH convergence}
\label{app:LiHconv}

\begin{figure*}
    \centering
    \includegraphics[width=1\linewidth]{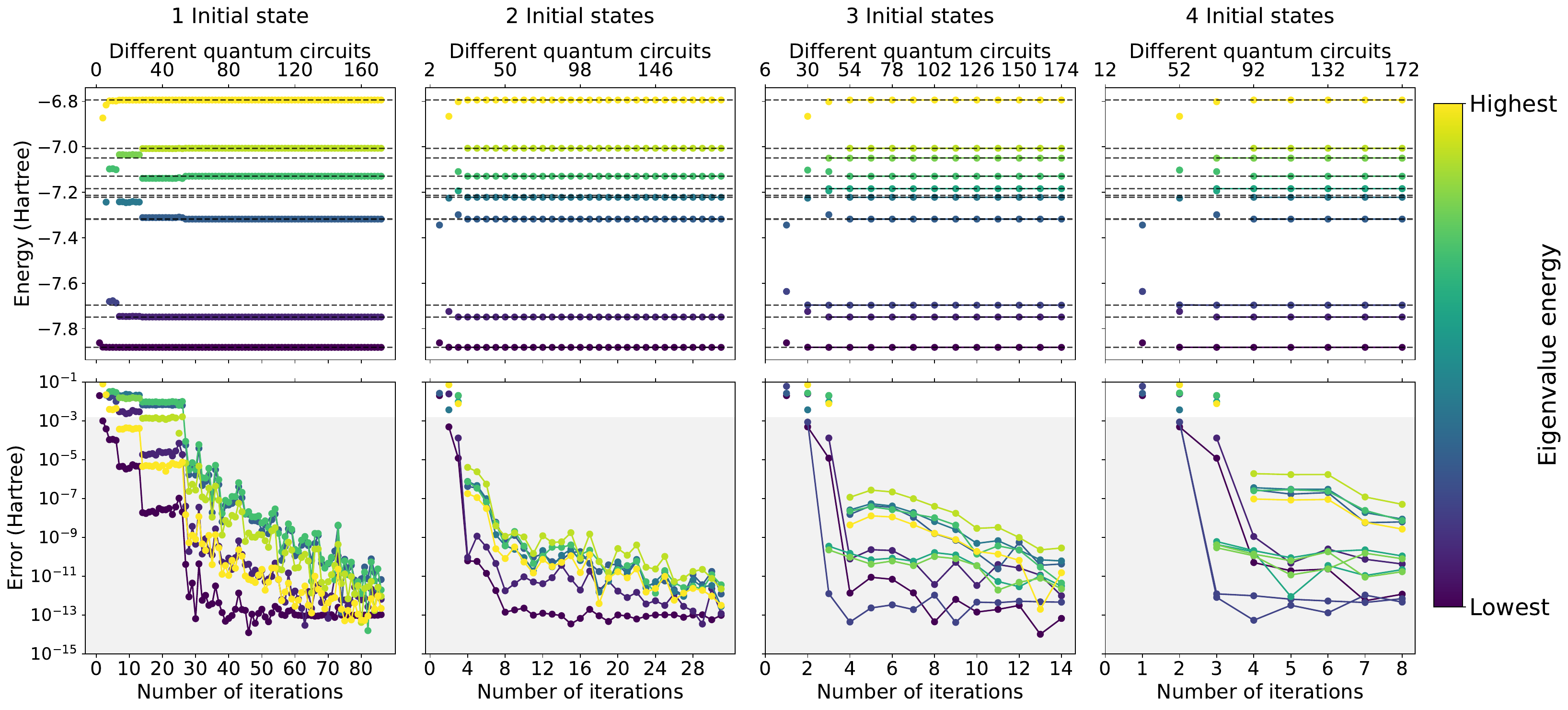}
    \caption{QBKSP algorithm applied to the LiH molecule for different numbers of initial states.
    The top panel shows the convergence of all eigenvalues both as functions of the number of iterations and the number of distinct quantum circuits required,  while the bottom panel shows the corresponding absolute errors as function of the same quantities.
    The grey region denotes values below chemical accuracy.
    }
    \label{fig:LIHinitial}
\end{figure*}

\begin{figure*}
    \centering
    \includegraphics[width=0.7\linewidth]{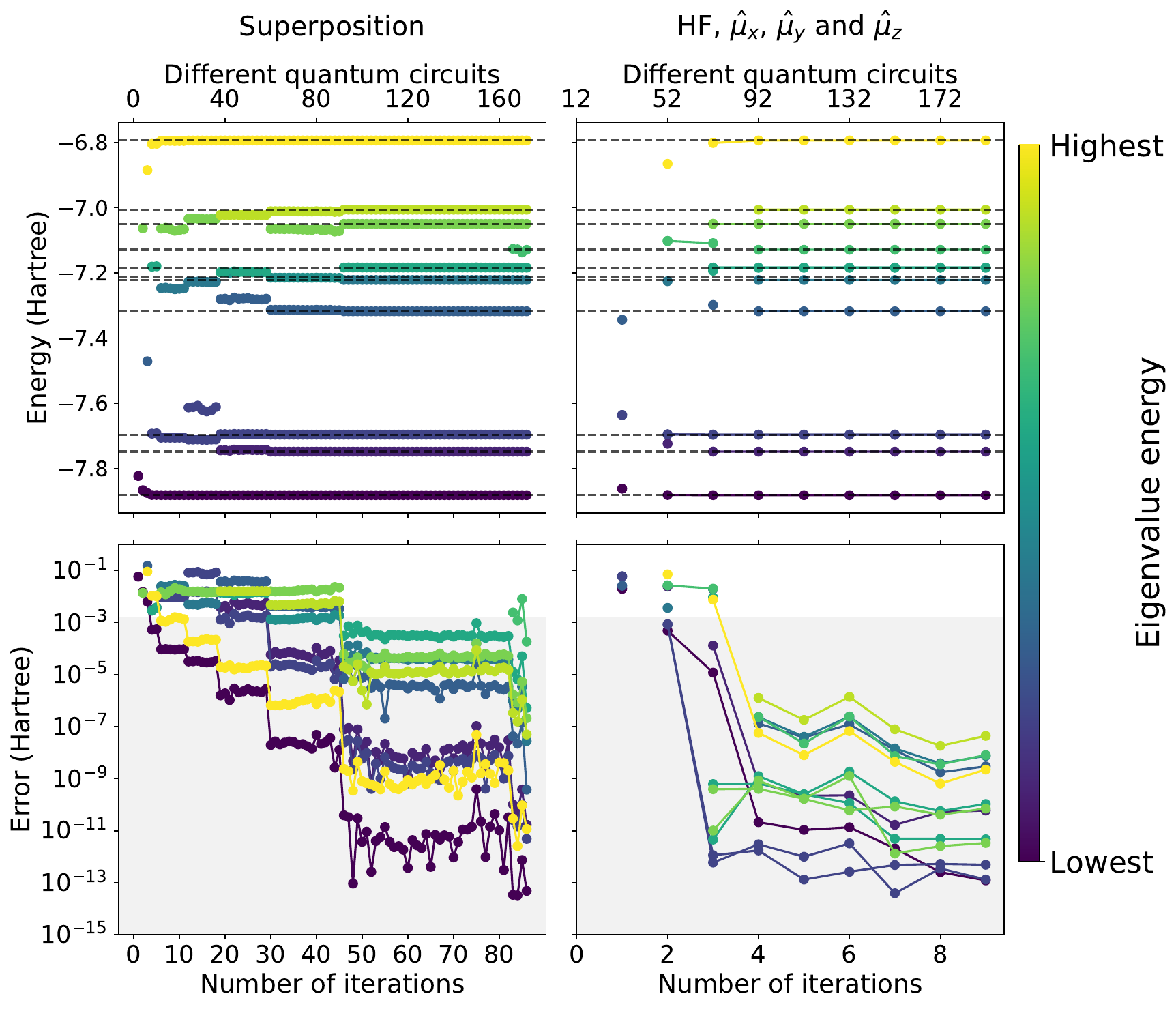}
    \caption{QBKSP algorithm applied to the LiH molecule for both one initial state corresponding to a uniform superposition of the HF state and the three dipole-excited variants of the HF and four initial reference states corresponding to these four states individually.
    The top panel shows the convergence of all eigenvalues both as functions of the number of iterations and the number of distinct quantum circuits required,  while the bottom panel shows the corresponding absolute errors as a function of the same quantities.
    The grey region denotes errors below chemical accuracy.
    }
    \label{fig:LiHsupvs4}
\end{figure*}

\autoref{fig:LIHinitial} shows the QBKSP convergence for the LiH molecule for different numbers of initial states and \autoref{fig:LiHsupvs4} shows the convergence for one single reference state corresponding to a uniform superposition of the HF state and the three dipole-excited variants of the HF and four initial reference states corresponding to these four states individually.

\section{Choice of reference states for molecular diatomics}
\label{app:initmole}

\begin{figure*}
    \centering
    \includegraphics[width=1\linewidth]{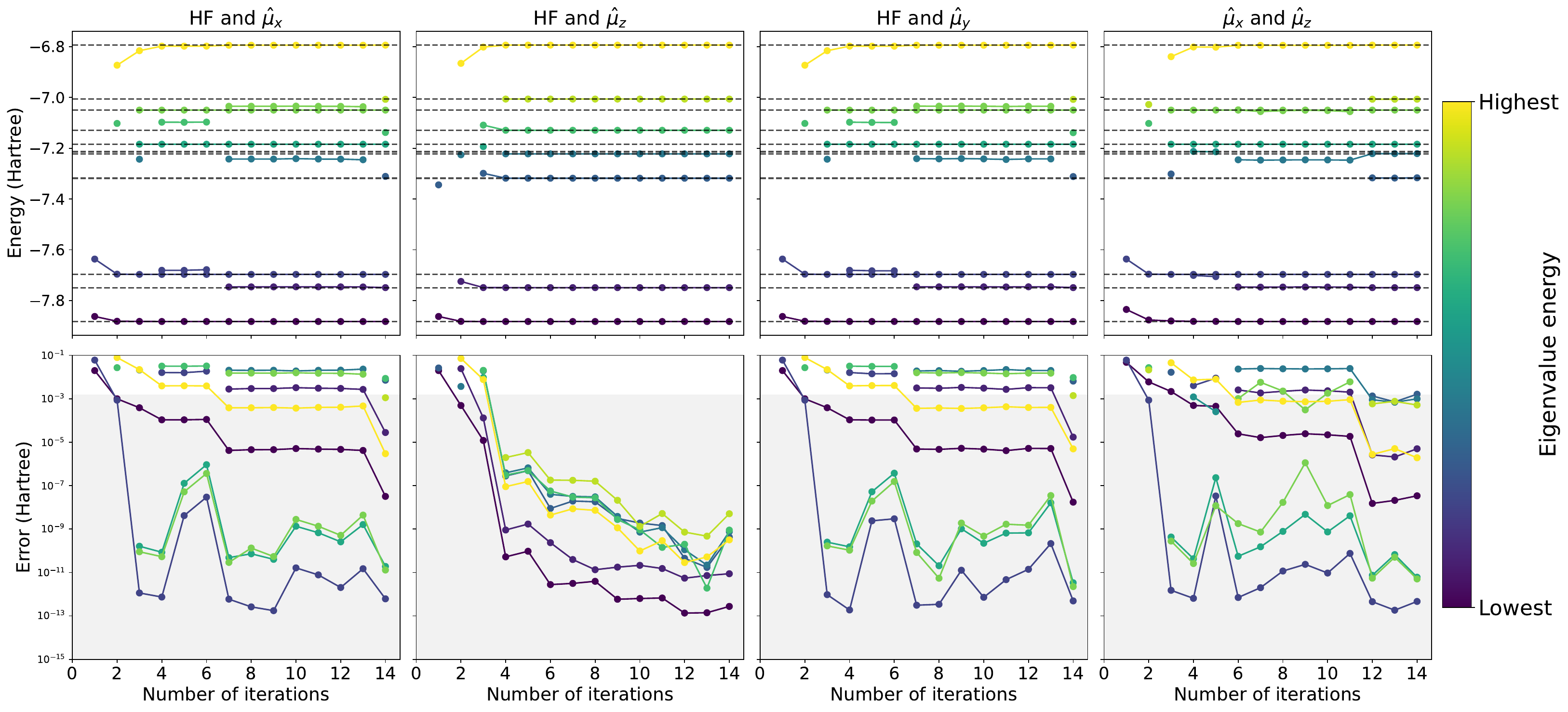}
    \caption{QBKSP convergence for the LiH molecule, evaluated for different combinations of 2 initial states.
    The top panel shows the convergence of all eigenvalues as a function of the number of iterations, whereas the bottom panel shows the absolute error with respect to the exact eigenvalues.
    }
    \label{fig:lih2}
\end{figure*}

\begin{figure*}
    \centering
    \includegraphics[width=1\linewidth]{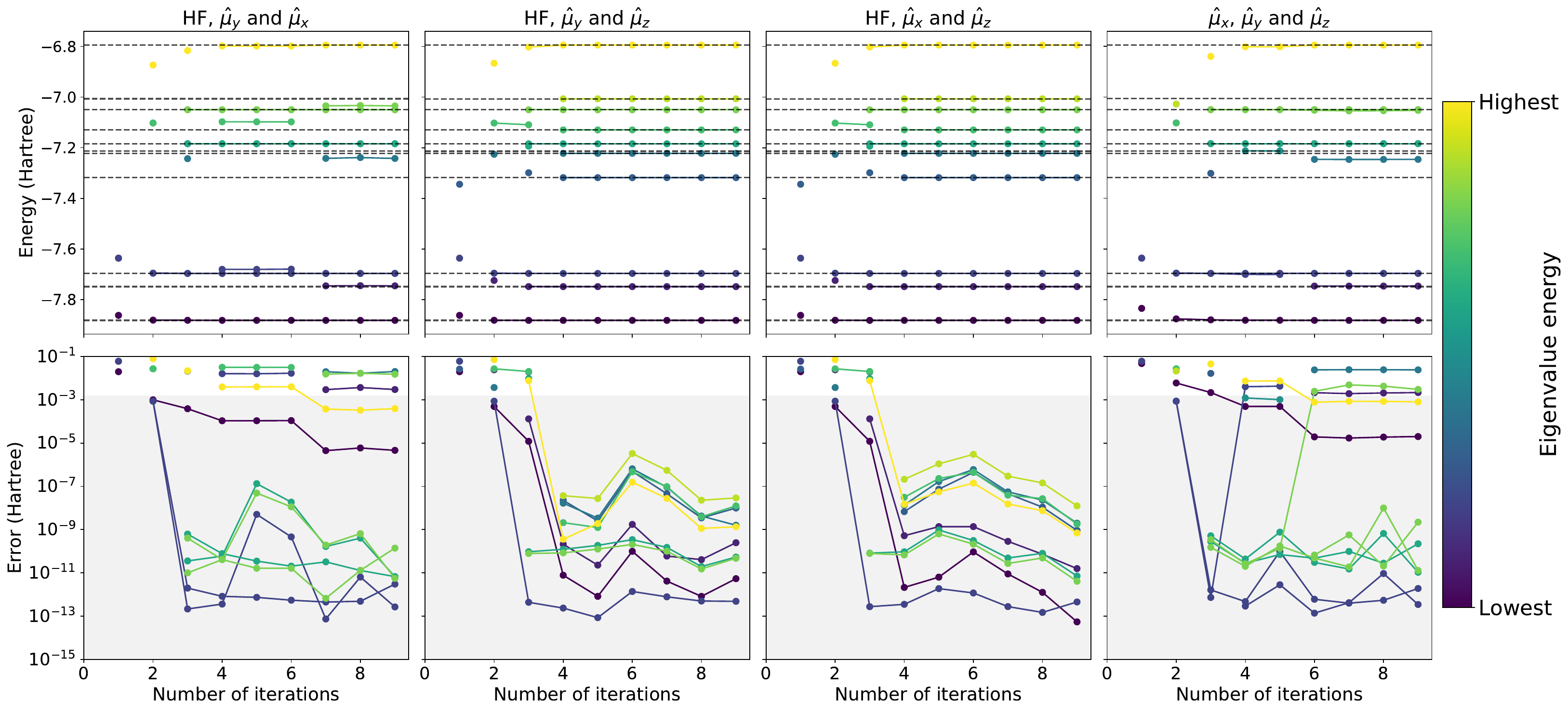}
    \caption{QBKSP convergence for the LiH molecule, evaluated for different combinations of 3 initial states.
    The top panel shows the convergence of all eigenvalues as a function of the number of iterations, whereas the bottom panel shows the absolute error with respect to the exact eigenvalues.
    }
    \label{fig:lih3}
\end{figure*}

To determine the optimal combination of two and three initial reference states out of the four states employed for the molecular diatomics investigated in this work, we apply the QBKSP algorithm to various combinations, as shown in \autoref{fig:lih2} for two references and \autoref{fig:lih3} for three references, using the LiH molecule as an example. 
The results clearly indicate that the optimal pair of reference states in this case is the HF state and the $\hat{\mu}_z$-excited state, while the optimal set of three references consists of the HF state and the $\hat{\mu}_y$- and $\hat{\mu}_z$-excited states.

\section{Trotter step size}
\label{app:trotter}

\begin{figure*}
    \centering
    \includegraphics[width=0.8\linewidth]{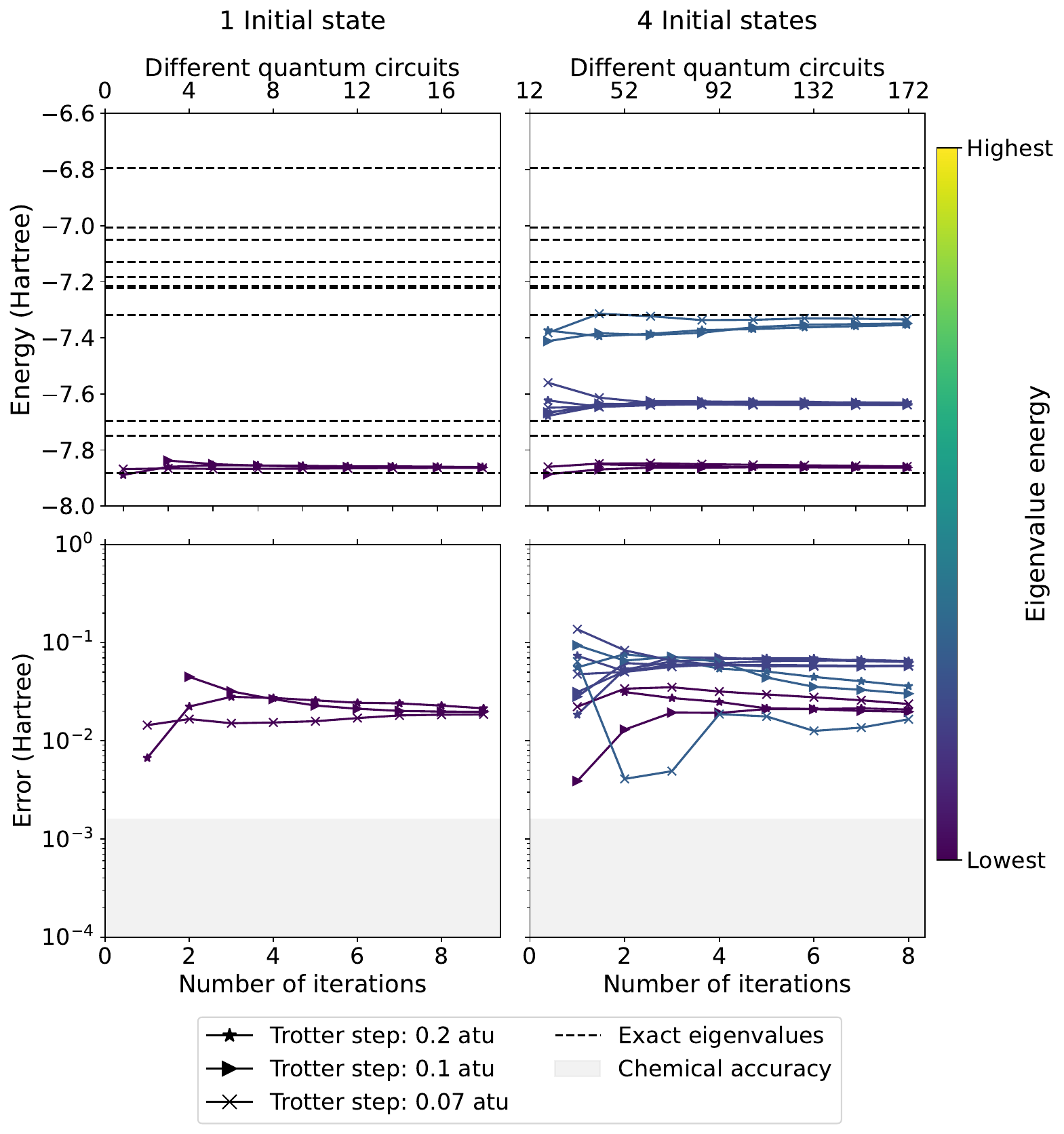}
    \caption{
    QBKSP algorithm executed on Qiskit \textit{Qasm Simulator} applied to the LiH molecule for different Trotter step sizes.
    The top panel shows the convergence of the obtained energies as a function of the Krylov iteration, whereas the bottom panel displays the absolute error with respect to the exact eigenvalues.
    }
    \label{fig:LiHtrotter}
\end{figure*}

We perform quantum simulations of the LiH molecule for both one reference state (the HF state) and four reference states (the HF and its three dipole-excited variants) using Qiskit \cite{qiskit2024} for different Trotter step sizes.
In \autoref{fig:LiHtrotter} the convergence of the eigenenergies is examined for three different Trotter step sizes for the same total evolution time of $\tau=1$ atu.
The $S$ matrix threshold parameter is set to $0.1$, and the number of shots is $10^5$.

\end{document}